\documentclass[fleqn,usenatbib]{mnras}

\usepackage{newtxtext,newtxmath}

\usepackage[T1]{fontenc}

\DeclareRobustCommand{\VAN}[3]{#2}
\let\VANthebibliography\thebibliography
\def\thebibliography{\DeclareRobustCommand{\VAN}[3]{##3}\VANthebibliography}

\usepackage{graphicx}	
\usepackage{amsmath}	
\usepackage{orcidlink}
\usepackage{bm}		
\usepackage{bookmark}
\usepackage{xcolor}  

\newcommand{\lx}{$L_{\text{X}}$}
\newcommand{\luv}{$L_{2500}$} 
\newcommand{\aox}{$\alpha_{\text{ox}}$}

\defcitealias{lamassa_31_2016}{LaM16}
\newcommand{\lamassa}{\citetalias{lamassa_31_2016}}

\include{Figs}

\DeclareRobustCommand{\vect}[1]{\bm{#1}}
\newcommand\hl[1]{{\textcolor{red}{#1}}}
\renewcommand{\hl}[1]{#1}  

\newcommand\hlii[1]{{\textcolor{red}{#1}}}
\renewcommand{\hlii}[1]{#1}

\title[Intrinsic X-ray luminosity distribution]{The intrinsic X-ray luminosity distribution of an optically-selected SDSS quasar population} 

\author[A. L. Rankine et al.]{
Amy L. Rankine$^{1}$\thanks{E-mail: amy.rankine@ed.ac.uk (ALR)}\orcidlink{0000-0002-2091-1966},
James Aird$^{1}$\orcidlink{0000-0003-1908-8463},
Angel Ruiz$^{2}$\orcidlink{0000-0002-3352-4383},
and Antonis Georgakakis$^{2}$\orcidlink{0000-0002-3514-2442}
\\
$^{1}$Institute for Astronomy, University of Edinburgh, Royal Observatory, Blackford Hill, Edinburgh EH9 3HJ, UK \\
$^{2}$Institute for Astronomy \& Astrophysics, National Observatory of Athens, V. Paulou \& I. Metaxa 11532, Greece
}

\date{Accepted XXX. Received YYY; in original form ZZZ}

\pubyear{2023}

\begin{document}
\label{firstpage}
\pagerange{\pageref{firstpage}--\pageref{lastpage}}
\maketitle

\begin{abstract}
In active galactic nuclei, the relationship between UV and X-ray luminosity is well studied (often characterised by {\aox}) but often with heterogeneous samples. We have parametrized the intrinsic distribution of X-ray luminosity, {\lx}, for the optically-selected sample of SDSS quasars in the Stripe 82 and XXL fields across redshifts 0.5--3.5. We make use of the available XMM observations and a custom pipeline to produce Bayesian sensitivity curves that are used to derive the intrinsic X-ray distribution in a hierarchical Bayesian framework. We find that the X-ray luminosity distribution is well described by a Gaussian function in ${\log_{10}}L_\text{X}$ space with a mean that is dependent on the monochromatic 2500\,{\AA} UV luminosity, {\luv}. We also observe some redshift dependence of the distribution. The mean of the {\lx} distribution increases with redshift while the width decreases. This weak but significant redshift dependence leads to {\luv}--{\lx} and {\luv}--{\aox} relations that evolve with redshift, and we produce a redshift- and {\luv}-dependent {\aox} equation. \hlii{Neither black hole mass nor Eddington ratio appear to be potential drivers of the redshift evolution.}
\end{abstract}




\section{Introduction}
\label{sec:intro}

The energetic processes associated with the fuelling of Active Galactic Nuclei (AGN) produce radiation across the electromagnetic spectrum. An optically thick accretion disc is expected to emit thermally resulting in a blackbody across the optical/UV \citep{shakura_black_1973}. Meanwhile, the bulk of the X-ray emission is thought to be produced by inverse Compton scattering of accretion disc photons accelerated to X-ray energies in some form of corona following a power-law spectrum. The geometry of the corona is unclear. Various models exist to describe this corona: from a lamp-post geometry where the corona illuminates the disc from its position above the black hole \citep{fabian_properties_2017}, to a slab corona that sandwiches the disc \citep{haardt_two-phase_1991}. \hl{X-ray polarimetry with the Imaging X-ray Polarimetry Explorer \citep[IXPE;][]{Weisskopf_2022_IXPE} has begun to constrain the geometry of the corona in a handful of AGN (MCG-05-23-16 [\citealt{marinucci_2022_polarization, tagliacozzo_2023_geometry}], IC 4329A [\citealt{ingram_2023_x-ray}], NGC 4151 [\citealt{gianolli_2023_uncovering}]).} The presence of a 'soft excess' (i.e., X-ray emission $\lesssim$1\,keV exceeding what would be expected from an extrapolated power-law) suggests an additional component to the X-ray production and is often attributed to an inner warm disc \citep{petrucci_testing_2018, petrucci_radiation_2020}.

The relationship between the X-ray and UV luminosity has been known for some decades \citep{avni_cosmological_1982, avni_x-ray_1986} and parametrized as $L_\text{X}\propto L_\text{UV}^\gamma$ with $\gamma\sim0.6$.
The relationship is generally considered to be tight \citep{lusso_quasars_2017, bisogni_chandra_2021}, although some scatter is observed, motivating models where the processes involved in producing the X-ray and UV emission are dependent on some common parameter of the AGN \citep[e.g., accretion rate, black hole mass;][]{lusso_quasars_2017, kubota_physical_2018}. 
There is little evidence of evolution with redshift \citep[e.g.,][]{vignali_x-ray_2003, steffen_x-ray--optical_2006, just_xray_2007, green_full_2009, lusso_quasars_2017, timliniii_what_2021}; however, see \hl{\citet{shen_soft_2006} and} \citet{kelly_evolution_2007} who do see some redshift-dependence of the relation.
In fact, the lack of significant redshift evolution and the general tightness of the relation has lead to claims that the correlation between the X-ray and UV luminosities (or more precisely the X-ray and UV fluxes) can be used to infer cosmological parameters \citep[][]{salvestrini_quasars_2019, lusso_quasars_2020}.

The spectral index of a power-law between the UV luminosity and X-ray luminosity, specifically the monochromatic 2500\,{\AA} and 2\,keV luminosities \citep[first introduced by][]{tananbaum_x-ray_1979}, is denoted {\aox} and is often used to parametrize the {\luv}--{\lx} relationship. \hl{\citet{jin_wavelength_2023} have recently shown that {\luv} is appropriate as a tracer of the accretion disc emission and that it is sufficient as a single parameter to describe the optical/UV emission over a broader wavelength range such that {\luv} is suitable for determining the relation of the UV to X-ray emission.} The non-flat relationship between {\aox} and {\luv} shows that the X-ray luminosity increases less than monotonically as UV luminosity increases suggesting that the spectral energy distribution (SED) becomes more disc-dominated. The physical driver of this relation is unclear and revealing the true relation between {\luv} and {\lx} free from selection effects will be a step towards understanding the physical mechanism(s) that govern the relations.

The observational {\luv}--{\lx} and {\luv}--{\aox} relations are plagued by selection effects due to both the choice of the parent quasar sample and the limitations of the available X-ray data. Many studies make attempts to reduce the systematic biases that can be introduced. In particular, in a flux-limited sample the correlation between luminosity and $z$ will invariably produce a redshift-dependent {\aox}. Attempts to reduce this effect include studying just the most luminous of sources across a wide redshift range ($z\approx1.5$--$4.5$) with the downside that the sample sizes are small \citep{just_xray_2007}; and adding a handful of faint $z\sim4$ AGN in order to remove the strong {\luv}--$z$ correlation \citep{kelly_evolution_2007}. While not making these particular choices at the sample selection stage, other studies have looked at the observed relations across narrow luminosity bins to determine the extent of any redshift evolution \citep{vignali_x-ray_2003}. Additionally, the X-ray non-detections must be treated with care. \citet{vignali_x-ray_2003, timliniii_what_2021} include upper X-ray flux limits for their X-ray undetected quasars. \citet{green_full_2009} do also but down-weight the undetected objects in their analyses. \citet{steffen_x-ray--optical_2006} include (optically-selected) objects with targeted X-ray observations such that the fraction of sources requiring upper X-ray flux limits is low. Meanwhile, \citet{lusso_quasars_2017} limit their sample to only sources that have X-ray detections.

In this paper, we develop and apply a Bayesian method to measure the intrinsic distribution of X-ray luminosities as a function of redshift and {\luv} for the well-defined sample of optically-selected SDSS quasars, carefully considering the impact of X-ray flux limits. We will make use of XMM observations in the Stripe 82 and XXL fields, reducing all of the XMM data with a custom pipeline in order to accurately construct the sensitivity curves in a consistent manner. Our approach is designed to not only use the X-ray detected quasar population but also extract information from X-ray undetected quasars with X-ray emission within the noise of the available XMM-Newton observations. With accurate sensitivity curves we will be able to consider the {\luv}--{\lx} relation in a probabilistic way, thereby removing the need for upper X-ray flux limits. The choice of using the optically-selected SDSS quasar population will reduce biases otherwise brought about by the inclusion of X-ray selected or radio-selected objects, for example. We will also then be able to produce {\luv}--{\lx} and {\luv}--{\aox} relations for a well-studied population of quasars and accurately determine any dependence on redshift.

We detail our sample selection criteria for the optically-selected sample and the careful reduction of the X-ray data, subsequent crossmatching, and calculations of luminosities in Section~\ref{sec:data}. In Section~\ref{sec:method} we describe our Bayesian methodology for calculating the underlying distribution of X-ray luminosity as a function of UV luminosity and redshift. The intrinsic {\luv}--{\lx} relation produced by our best-fitting model is presented in Section~\ref{sec:LuvLx} followed by the corrected {\luv}--{\aox} relation in Section~\ref{sec:aox}. We briefly discuss our finding of an evolving {\luv}--{\lx} relation with redshift in Section~\ref{sec:disc}.

Vacuum wavelengths are employed throughout the paper and we adopt a $\Lambda$CDM cosmology with $h_0 = 0.71$, $\Omega_\text{M} = 0.27$, and $\Omega_\Lambda = 0.73$ when calculating quantities such as quasar luminosities.

\section{Data}
\label{sec:data}
We use X-ray data from XMM and UV/optical data from SDSS both taken in the XXL and Stripe 82 fields. In brief, we are using the optically-selected quasars from SDSS DR16 \citep{lyke_sloan_2020} at redshifts $0.5 < z < 3.5$ across the two regions and have re-reduced the XMM data using the {\sc xmmpype} custom pipeline outlined in \citet{georgakakis_serendipitous_2011}. We crossmatch the XMM sources with the SDSS sources using {\sc Nway} \citep{salvato_finding_2018}. Detailed descriptions of each dataset are provided below; however, some readers may wish to peruse Table~\ref{tab:sample} and move onto Section~\ref{sec:method}.

\subsection{Optical/UV}
\label{ssec:opt}

We use the SDSS DR16 quasar catalogue \citep{lyke_sloan_2020} to create the optically-selected sample of AGN within the XXL and S82 fields. We filter the DR16 quasar catalogue with the multi-order coverage maps (MOCs) of the XXL and S82 XMM observations (see Section~\ref{ssec:xray}) in {\sc Aladin} to select only the objects within SDSS that fall within the footprints of XXL and S82. The quasar catalogue is further limited to the optically-selected quasars which we define as the CORE sample from the BOSS and eBOSS targets \citep{myers_sdss-iv_2015}. The CORE sample is produced by selecting objects with the following of SDSS's Bitmasks activated: bit 40 (\verb|QSO_CORE_MAIN|) of mask \verb|BOSS_TARGET1|, bit 10 (\verb|QSO_EBOSS_CORE|) of \verb|EBOSS_TARGET0|, and bit 40 (\verb|QSO1_EBOSS_CORE|) of \verb|EBOSS_TARGET1|. With this selection, we aim to only include quasars that were selected and targeted based on their optical properties. In doing so, we avoid biasing our results by including, for example, the X-ray selected quasars in the XXL field which were observed as part of the large SDSS ancillary programme led by A. Georgakakis.

The optical sample contains both X-ray detected and undetected objects (see Section~\ref{ssec:xray}) with a total of 2292 quasars. \hl{Note that our selection does not remove quasars with broad absorption lines in their spectra \citep[BAL quasars;][]{weymann_1991_comparisons} or radio-loud (i.e., jetted) quasars \citep{kellermann_1989_vla}. This choice was made in order to assess the X-ray properties of the truly \textit{optically}-selected quasar population; removing them would introduce additional selection biases. However, we acknowledge that BAL quasars tend to be X-ray weak compared to non-BAL quasars \citep[e.g., ][]{gibson_2009_catalog,luo_2014_weak} and the X-ray emission of radio-loud quasars can be dominated by jets \citep{shang_2011_next, zhou_2021_composite}. BAL quasars can only be identified at $z>1.5$ at which redshifts any potential \ion{C}{iv} absorption systems are within the observed wavelength window of SDSS, meaning that we are only able to identify BAL quasars in 54\,\% of our sample. There are only 58 quasars identified as BAL quasars in our optical sample, of which 5 are X-ray detected. Radio-loud quasars number only 32 in the sample \citep[based on having a match to FIRST in the SDSS DR16 quasar catalogue;]{lyke_sloan_2020}, with 17 of these detected in the X-ray. It is unknown if there are additional radio-loud quasars in our sample that would be detected with deeper data, further justifying our decision to not apply a radio cut on our sample.}

\subsubsection{Optical/UV properties}

\begin{figure*}
    \centering
    \includegraphics[width=\linewidth]{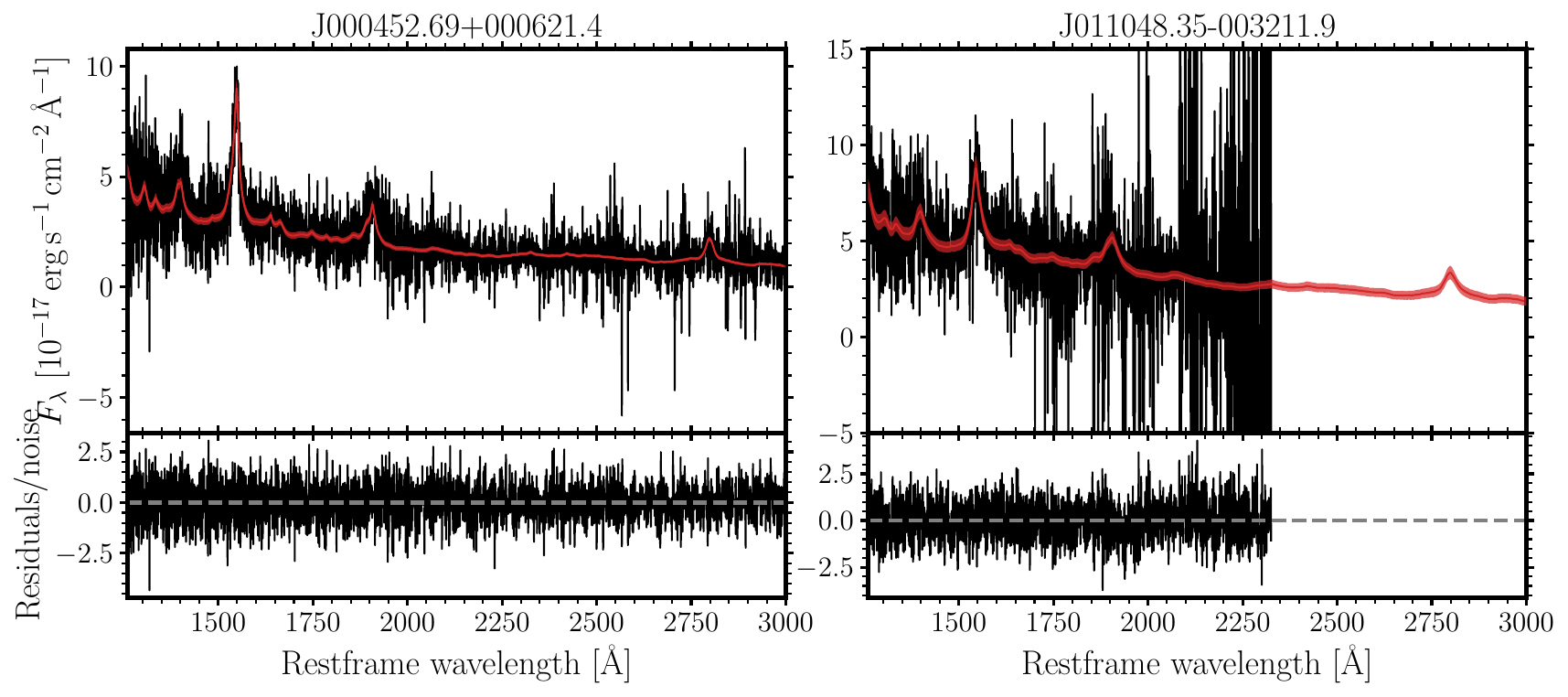}
    \caption{Example reconstructions of quasar spectra using the ICA technique for restframe 1260--3000\,{\AA} (top) and the residuals (observed spectrum flux $-$ reconstruction) normalised by the noise (bottom). The left panels contain a $z=2.08$ quasar with full coverage of the 2500\,{\AA} region. The right panels demonstrate the extrapolation of the reconstruction for a $z=3.47$ quasar without coverage of the 2500\,{\AA} region. The red shaded area is the 1-$\sigma$ uncertainties on the reconstruction.}
    \label{fig:spec}
\end{figure*}

The SDSS spectra are reconstructed using the ICA technique outlined in \citet{rankine_bal_2020} which essentially provides high S/N versions of the spectra over the restframe wavelength range 1260--3000\,{\AA} from which the continuum luminosity at restframe 2500\,{\AA} can be measured, {\luv}. The left-hand panel of Fig.~\ref{fig:spec} contains an example spectrum and reconstruction. {\luv} is estimated by calculating the median flux in a 10\,{\AA} window centred on 2500\,{\AA} and converting to a luminosity. \hl{Where available, redshifts from \citet{rankine_bal_2020} which are based on an independent component analysis (ICA) of the optical spectra are used, otherwise, the redshifts reported in \citet{lyke_sloan_2020} are used. The differences between the two redshift samples are of order 300\,km\,s$^{-1}$ with only a few as different as $\sim$1000\,km\,s$^{-1}$. The updated redshifts from \citet{rankine_bal_2020} will not significantly affect the calculations of luminosities; however, they will produce more accurate black hole mass measurements, particularly \ion{C}{iv}-based masses due to the correction derived from the `blueshift' of the emission line (see Section~\ref{sec:disc}). All in all, the changes are minimal.}

We correct the luminosities for Galactic dust extinction with the {\sc dustmaps} Python module \citep{green_dustmaps_2018} and the dust map of \cite{Schlegel_maps_1998} updated by \citet{schlafly_measuring_2011} in tandem with the {\sc extinction} module \citep{barbary_extinction_2016} and the reddening curve of \citet{fitzpatrick_correcting_1999}, producing median $E(B-V)=0.02$ for the XXL sample and 0.03 for S82 and S82X. 2500\,{\AA} is redshifted out of the BOSS spectrograph at $z\gtrsim3.2$ ($z\gtrsim2.7$ for the SDSS spectrograph) which would ordinarily prevent the measurement of the 2500\,{\AA} monochromatic luminosity of quasars above this redshift. However, reconstructing the spectra with the ICA technique which utilises the spectral information, including emission lines and the continuum shape, across the rest of the available spectrum above 1260\,{\AA} allows the 2500\,{\AA} luminosity to be estimated reliably. We checked the accuracy of extrapolating the reconstructions with a sample of quasar spectra in which 2500\,{\AA} was present but only included the wavelength range 1260--2200\,{\AA} in the fitting and found good agreement with the reconstructions that used the full available wavelength range between 1260--3000\,{\AA}. See an example of this extrapolation in the right-hand panel of Fig.~\ref{fig:spec}. Only 46 quasars of our optically-selected sample require extrapolation of the reconstructions.

Uncertainties on $L_{2500}$ are calculated by propagating the errors on the weights of the ICA spectral components produced during the reconstruction process. The median errors on $L_{2500}$ for the subset of objects without restframe 2500\,{\AA} in their spectra and {\luv} was extrapolated from the reconstructions are $\sim$0.04\,dex compared to $\sim$0.02\,dex for the subset with restframe 2500\,{\AA} which reflects the indirect measurement of $L_{2500}$. Errors from the spectrum reconstructions will be much less than those from the spectrophotometry; however, we do not propagate the {\luv} errors further, since the main source of uncertainty is the X-ray luminosities, and so do not make an attempt to quantify them here.

\begin{figure*}
    \centering
    \includegraphics[width=\linewidth]{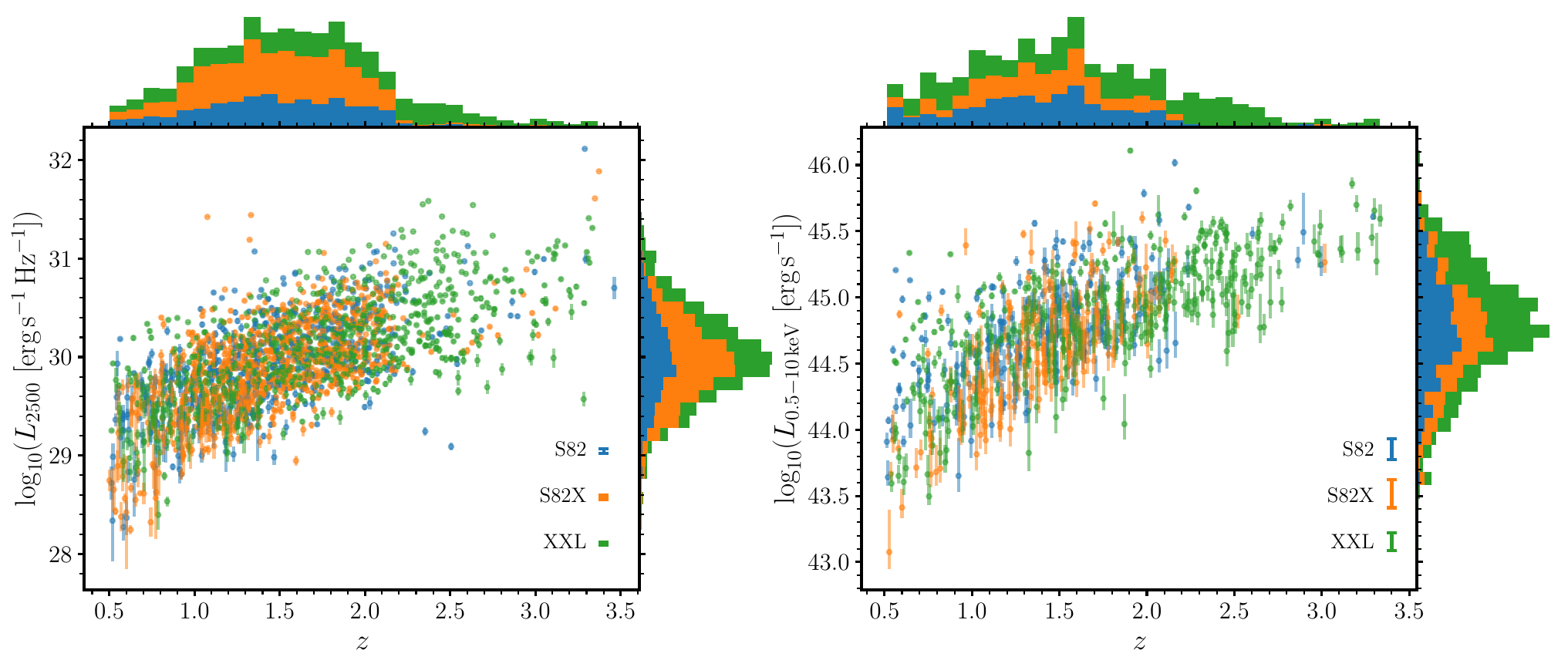}
    \caption{Optical/UV (left) and X-ray luminosities (right) as a function of redshift for the Stripe 82 (blue), Stripe 82X (orange) and XXL (green) samples. Median luminosity errors are presented in the legend. The 1-D redshift and luminosity distributions for the three samples are plotted above and to the right, respectively, of their corresponding axes. The {\luv} panel contains both X-ray detected and undetected quasars, whereas the {\lx} panel contains only the X-ray detected subsample.}
    \label{fig:luvlxz}
\end{figure*}

The left panel of Fig.~\ref{fig:luvlxz} shows the distribution of $L_{2500}$ versus redshift for the X-ray detected and undetected quasars. The CORE Stripe 82 and Stripe 82X samples contain very few quasars above $z\sim2.2$ compared to the XXL sample due to the differing SDSS selection between SDSS II and SDSS III/IV with all of the CORE Stripe and Stripe 82X quasars originating from SDSS II.

\subsection{X-ray}
\label{ssec:xray}
We start from the 294 XMM pointings in the North field of XXL \citep{pierre_xxl_2016}. XMM-XXL North covers $\sim$25\,deg$^2$ with an exposure time of 10 ks per XMM pointing.

Stripe 82 is an equatorial region of sky covering $\sim$300\,deg$^2$ which has been repeatedly observed with SDSS. Approximately 28\,deg$^2$ of Stripe 82 has been observed with XMM. This combines the 198 pointings from the Stripe 82X survey (S82X) at $\sim$5\,ks per XMM pointing and 33 additional archival pointings (S82; 7-66\,ks per pointing) extracted from the XMM archive \citep{lamassa_finding_2013b, lamassa_31_2016}.

\subsubsection{Reduction}

\begin{figure}
    \centering
    \includegraphics[width=\linewidth]{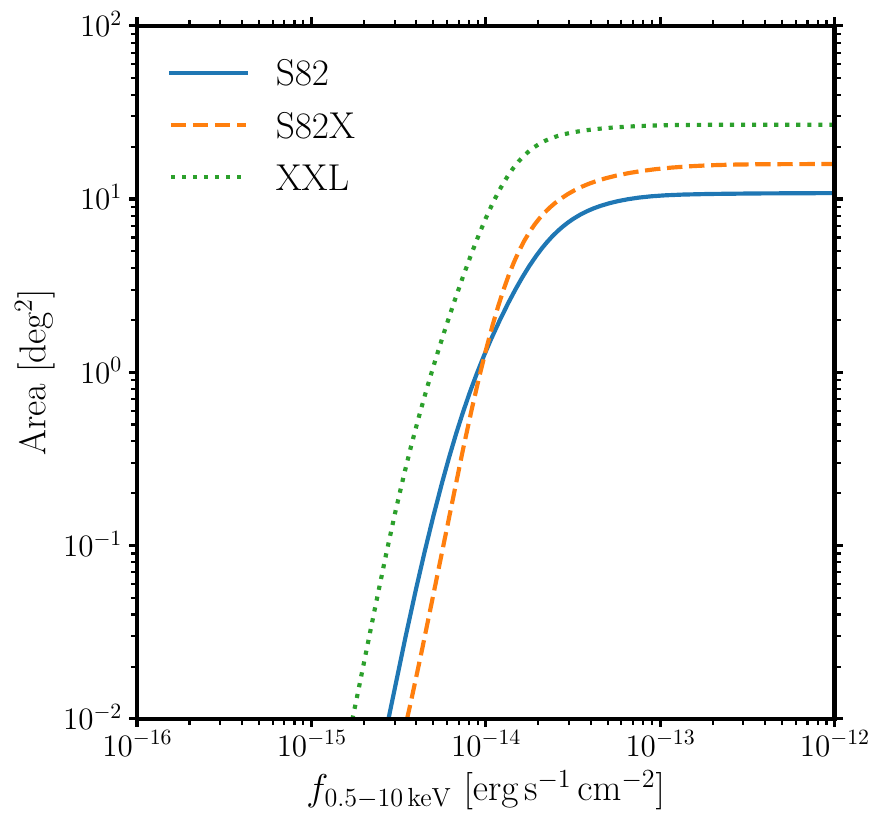}
    \caption{Sensitivity curves for the full band (0.5--10\,keV) across the three regions in our sample.}
    \label{fig:acurve}
\end{figure}

\begin{figure}
    \centering
    \includegraphics[width=\linewidth]{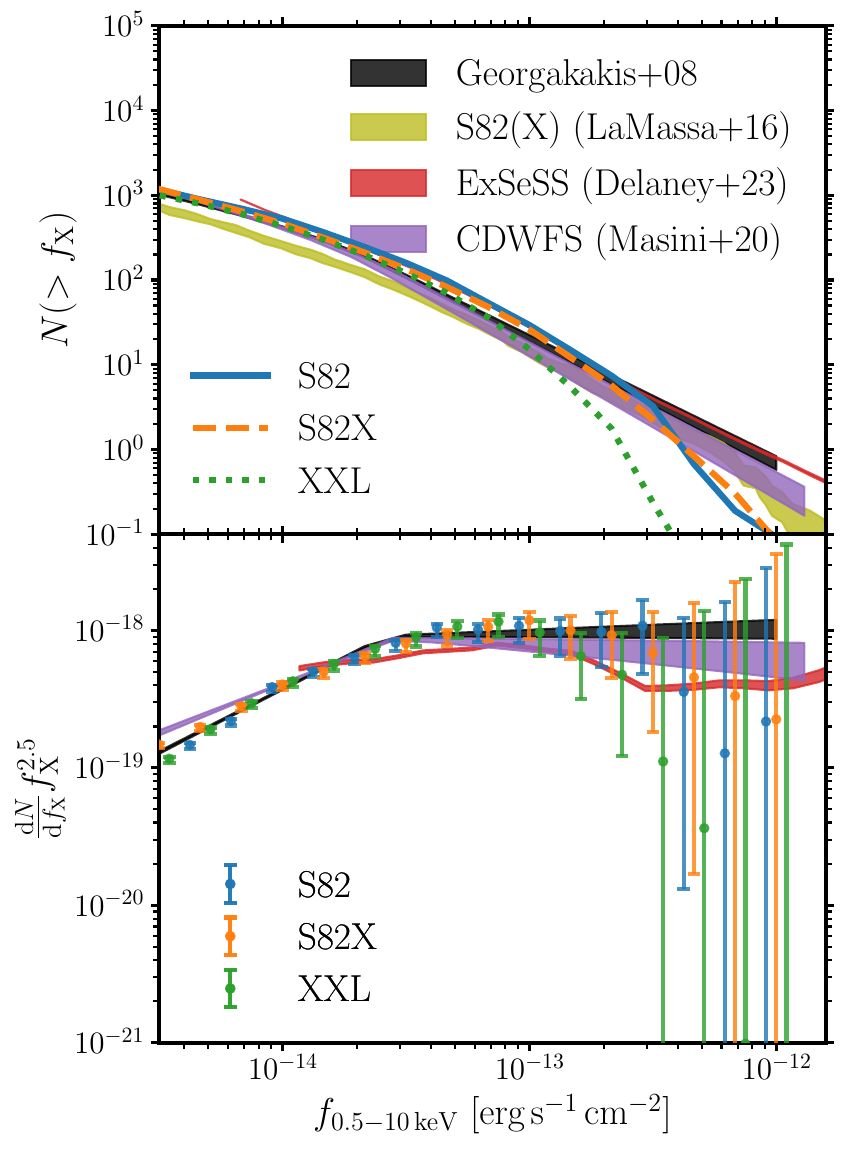}
    \caption{\hl{Top: cumulative number counts as a function of full band XMM flux and comparison to ExSeSS \citep{delaney_extragalactic_2023}, CDWFS \citep{masini_chandra_2020}, the reduction of Stripe 82 by \citet{lamassa_31_2016}, and the model of \citet{georgakakis_new_2008}. Bottom: differential number counts with the Euclidean slope removed. Errors are Poisson errors based on the number of sources and scaled accordingly. All measurements have been converted to the 0.5-10\,keV band assuming a consistent $\Gamma=1.4$ spectrum.}}
    \label{fig:logNlogS}
\end{figure}

We use the {\sc xmmpype} XMM pipeline, which is based on the methods and techniques described in \citet{georgakakis_serendipitous_2011}.
In brief, the pipeline creates images in the different energy bands, sources are detected and astrometric corrections are applied before X-ray fluxes are estimated and any optical counterparts to the X-ray sources are identified.
One advantage of employing the pipeline is the greater accuracy of the sensitivity curves which are generated with a robust and well-quantified Bayesian approach \citep[following the methods of ][]{georgakakis_new_2008}.
The sensitivity curves allow for an accurate characterisation of the selection function of a sample using analytic relations instead of cumbersome and computationally expensive simulations and can naturally account for non-detected sources. 
In particular, at faint fluxes the Bayesian sensitivity curves correctly account for the effects of Poisson statistics on the X-ray detection and photometry in the low-counts regime and the impact of Eddington bias. 
Figure~\ref{fig:acurve} contains the area curves for the S82, S82X, and XXL fields in the full band. In general, at a given flux, the XXL sample is most sensitive, followed by the S82 archival pointings and finally the S82X survey.
The nature of our investigations means that correcting for the X-ray detection probability is necessary and will be most significant at faint fluxes.
Additionally, the pipeline coadds overlapping XMM observations to increase the X-ray depth. It is also designed for large-area serendipitous X-ray surveys which greatly facilitates the post-processing of the various products in the case of surveys that extend over large sky areas.
We limit our sources to those detected in the full band (0.5--10\,keV) 
where a detection is defined by a ``false detection probability'' $p_\mathrm{false}<4\times10^{-6}$, where $p_\mathrm{false}$ is the probability of the observed counts (or higher) being produced purely by a fluctuation of the background.
Column 1 of Table~\ref{tab:sample} lists the number of X-ray point sources resulting from the reduction of the XMM pointings, totalling 14\,493 sources. Comparing to the S82 reductions of \citet{lamassa_31_2016}, we find 5529 X-ray sources in the combined S82 regions, whilst \citet{lamassa_31_2016} produced a catalogue of 4668 sources with XMM detections in the full band.
We find that the $\log N$--$\log S$ relations of \citet{georgakakis_new_2008}, the ExSeSS catalogue \citep{delaney_extragalactic_2023}, and the CDWFS \citep{masini_chandra_2020} are in good agreement with those of our sample (see Fig.~\ref{fig:logNlogS}) providing confidence in the source detection and sensitivity maps of the {\sc xmmpype} reductions (see Appendix~\ref{app:lamassa} for comparisons in the hard and soft bands).

\begin{table}
    \centering
    \caption{Number counts for final samples of X-ray detected and undetected sources for the S82, S82X and XXL fields. The first column contains the total number of point sources with detections in the full band extracted with {\sc xmmpype}. The second and third columns contains the number of optically-selected quasars that have X-ray counterparts (are X-ray detected) and those that do not (undetected).}
    \label{tab:sample}
    \begin{tabular}{cccc}
        \hline
            & Sources & Detected & Undetected \\
        \hline
        S82 & 2393 & 226 & 366 \\
        S82X & 3136 & 196 & 764 \\
        XXL & 8964 & 348 & 392 \\
        \hline
    \end{tabular}
\end{table}

\subsubsection{Crossmatching}
We perform an initial search for possible optical counterparts in SDSS DR16 \citep{ahumada_16th_2020} with {\sc xmatch} \citep{xmatch_2020} and a search radius of 40 arcsec around each X-ray source which yields an optical catalogue of 1\,484\,651 sources. We use {\sc Nway} to match the X-ray observations to this catalogue with a 20 arcsec maximum radius. X-ray RA and Dec positional uncertainties were generated during the reduction with median uncertainties of $\sim$1.5 arcsec. We supply constant 0.1 arcsec positional uncertainties for the optical catalogue. We supply {\sc Nway} with the total sky area of the reduced XMM observations -- calculated from the multi-order coverage maps (MOCs) generated by {\sc xmmpype} -- and estimate the sky area of the input optical catalogue by creating a MOC with {\sc Aladin} \citep{aladin_2000} and a radius around each X-ray observation of 40 arcsec, producing a total area of 14.23\,deg$^2$ once overlaps between 40-arcsec regions have been accounted for. {\sc Nway} produces a matched catalogue containing all possible matches for each X-ray source and corresponding probabilities. $p_{\text{any}}$ is the probability that an X-ray source has a true counterpart in the provided catalogue and $p_{i}$ is the probability that a particular match is the true counterpart. As such, a combination of $p_{\text{any}}$ and $p_{i}$ and limits on each can be invoked to produce a final catalogue of robust optical counterparts of the X-ray sources. 
{\sc Nway} calculates the average source density on the sky from the provided sky areas which leads to a scaling of the counterpart probabilities, $p_{\text{any}}$ and $p_{i}$. Combining the XXL and Stripe 82 fields leads to an average sky density across the two fields which will affect the relative counterpart probabilities for sources in different fields.
However, in our use case of {\sc Nway} we do not use any absolute $p_\text{any}$ or $p_i$ thresholds to determine the final matches; instead we are only ever comparing $p_\text{any}$ and $p_i$ values between different objects across small physical scales; i.e., optical sources that are potential matches to the same X-ray source such that they are within the same region. As such, the scaling of the probabilities does not affect the final matching.

We include magnitude priors in the crossmatching to preferentially select counterparts with \hl{optical magnitudes that match the magnitude distribution of quasars which are less likely to be spurious alignments and more likely to be the true counterparts to the X-ray sources. We perform the matching with $r$-band information from SDSS with priors pre-determined based on the magnitudes of the optical quasars in the optical catalogue compared to the non-quasar objects.}

We make the final match selection by prioritising counterparts that are classed as AGN which we define as either having \verb|spCl| (the spectroscopic class) as `QSO' or if the object is found in the SDSS DR16 quasar catalogue \citep{lyke_sloan_2020} having performed a simple 1-arcsec crossmatch between the DR16 and DR16Q catalogues. To implement the AGN-prioritisation, we take the possible matches from {\sc Nway}, and inspect the match with the highest product of $p_{\text{any}}$ and $p_i$ that is also an AGN. The product avoids multiple X-ray sources having the same AGN optical counterpart and gives priority to the X-ray source with the highest probability of having a counterpart in this optical catalogue ($p_{\text{any}}$). Only using $p_i$ would result in 195 X-ray sources having an optical match already associated with another X-ray source.
Only if the $p_i$ for this optical AGN is $>0.01 p_{i}$ of the original best match is the AGN selected as the counterpart.
We perform a false-positive calibration by offsetting the X-ray positions and running {\sc Nway} with this mock X-ray catalogue (and corresponding mock optical catalogue obtained with {\sc xmatch} and a 40 arcsec search radius).
The AGN number density on the sky is low such that for our AGN-prioritisation scheme a $p_{\text{any}}$ threshold of zero is sufficient to maintain a false-positive fraction $<$1\,\%. 

Ultimately, we obtain the highest completeness when including the $r$-band quasar-based magnitude prior; however, the QSO-prioritisation scheme leads to only a few X-ray matches changing depending on the prior used. \hl{We end up with 26\,\% of all our X-ray sources having an optical counterpart that is spectroscopically identified as an AGN.} Given that we are starting from an optically-selected subsample of the SDSS DR16 quasar catalogue, we limit the sample to the X-ray sources that have optical counterparts identified as AGN based on their inclusion in the DR16 quasar catalogue. Our final sample thus contains 2292 optically-selected AGN, 770 (34\,\%) of which are X-ray detected (see Table~\ref{tab:sample}).

\subsubsection{X-ray properties}

\begin{figure}
    \centering
    \includegraphics[width=\linewidth]{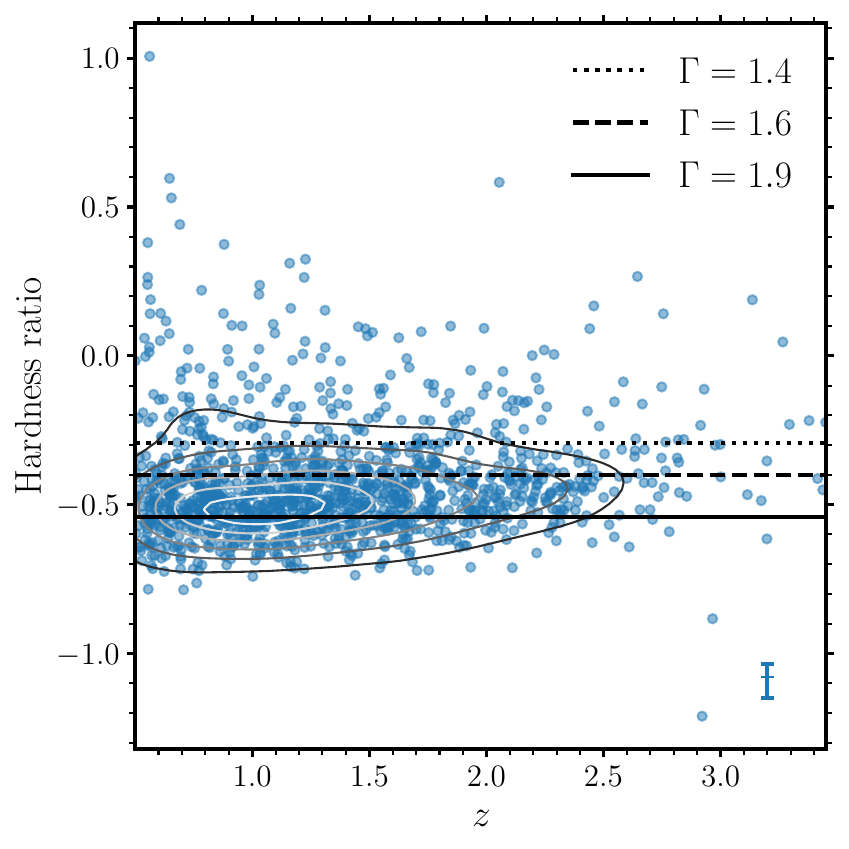}
    \caption{Hardness ratios versus redshift for our X-ray detected quasar sample. The median errors are plotted in the bottom right. The horizontal lines mark the average hardness ratios for $\Gamma=1.4$, 1.6, and 1.9 assuming appropriate \ion{H}{i} column densities of $2\times10^{20}$ and $3\times10^{20}$\,cm$^{-2}$ and PN and MOS detectors.}
    \label{fig:HR}
\end{figure}

X-ray flux measurements for the full 0.5--10\,keV band are calculated during the reduction with Galactic absorption taken into account \citep[estimated from the \ion{H}{i} maps of the LAB survey;][]{kalbera_leiden_2005} but assume a photon index of $\Gamma=1.4$. We are specifically selecting X-ray sources associated with (broad-line) quasars and so expect them to have unabsorbed X-ray spectra. \hl{We check this using the hardness ratios, defined as
\begin{equation}
    \text{HR} = \frac{H-S}{H+S},
\end{equation}
where $H$ and $S$ are source counts in the hard (2--10\,keV) and soft bands (0.5--2\,keV) normalised by exposure, and confirm that they are, on average, consistent with $\Gamma=1.9$} (see Fig.~\ref{fig:HR}). We convert the flux measurements to a $\Gamma=1.9$ using conversion factors from webpimms based on \ion{H}{i} column densities of $2\times10^{20}$ and $3\times10^{20}$\,cm$^{-2}$ for the XXL and S82 fields, respectively. We apply a K-correction when calculating rest-frame luminosities that also assumes a photon index of $\Gamma=1.9$.
The X-ray luminosity distribution with redshift is plotted in the right-hand panel of Fig.~\ref{fig:luvlxz}. The sensitivity of the X-ray data is apparent in the lower bound on {\lx} with redshift. The S82 and XXL samples are similar in their {\lx}--$z$ distributions; however, XXL extends to higher redshifts due to the SDSS selection.

\section[Measurements of the intrinsic distribution of \texorpdfstring{\lx}{Lx} as a function of \texorpdfstring{\luv}{L2500} and redshift]{Measurements of the intrinsic distribution of $\vect{L_\text{X}}$ as a function of $\vect{L_{2500}}$ and redshift}
\label{sec:method}

\begin{figure*}
    \centering
    \includegraphics[width=\linewidth]{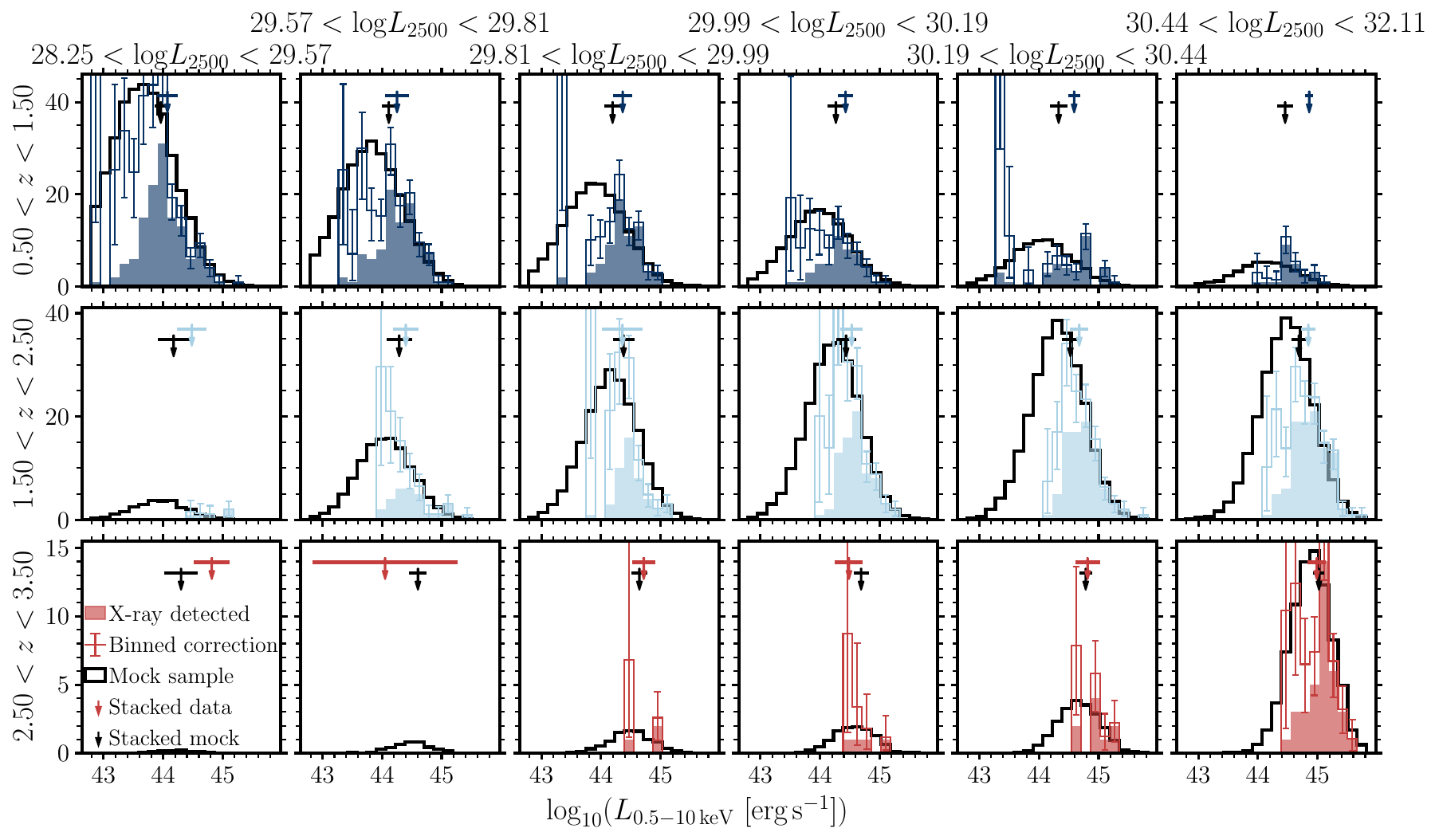}
    \caption{Distribution of {\lx} binned by {\luv}, $z$. Each panel is a different {\luv} (columns, increasing to right) and $z$ (rows, increasing towards bottom) bin. The X-ray detected quasars are presented as the filled histograms. The binned corrected counts and associated Poisson errors are represented by the open histograms and error bars (see Section~\ref{sec:binned}). The black histograms are a random sample drawn from the assumed Gaussian distribution of {\lx} with parameters determined by the maximum likelihood estimation with the best-fitting model (vii) (see Section~\ref{sec:mle}). From both the binned corrected counts and the MLE results, it is clear that the X-ray detected sample is skewed towards the high {\lx} sources. In the majority of the {\luv} and $z$ bins the stacked {\lx} from the MLE results agrees with the stacked data (black and coloured vertical arrows with 1-$\sigma$ error bars). Only the ({\luv}, $z$, {\lx}) bins populated with X-ray detected quasars can be corrected via the binning method outlined in Section~\ref{sec:binned}, providing motivation for the MLE detailed in Section~\ref{sec:mle}).}
    \label{fig:binnedLx}
\end{figure*}

We aim to arrive at a model that describes the distribution of X-ray luminosity as a function of UV luminosity and redshift. In Fig.~\ref{fig:binnedLx} we plot the distribution of X-ray luminosity for our X-ray detected quasar sample in bins of {\luv} and $z$ (solid colour histograms). In what follows we will make use of the Bayesian sensitivity curves provided by {\sc xmmpype} in order to account for the X-ray undetected quasar population and derive the underlying {\lx} distribution function. 

\subsection[Completeness-corrected distribution at a given \texorpdfstring{\lx}{Lx}]{Completeness-corrected distribution at a given $\vect{L_{\text{X}}}$}
\label{sec:binned}
We attempt to account for the undetected X-ray sources in each ({\lx}, {\luv}, $z$) bin by calculating the probability of a source having an X-ray luminosity {\lx} given its {\luv} and $z$:
\begin{equation}
    P(L_{\text{X}}|L_{2500}, z) = \frac{N_{\text{det}}}{\sum_{i=1}^{N_{\text{tot}}} p({\text{det}}|L_{{\text{X}}}, z_i)\,\Delta {\log_{10}}L_{\text{X}}}
\end{equation}
The numerator is the number of X-ray detected quasars in each ({\lx}, {\luv}, $z$) bin. The denominator takes into account the probability that quasar $i$ with redshift $z_i$ would be detected if it had an X-ray luminosity corresponding to the centre of the {\lx} bin and is summed over all X-ray detected and undetected quasars in that ({\luv}, $z$ bin). $\Delta {\log_{10}}L_{\text{X}}$ is the width of the {\lx} bin. The corrected counts in a given ({\lx}, {\luv}, $z$) bin can then be calculated by the following:
\begin{equation}
    \begin{split}
        N_{\text{corr}} &= P(L_{\text{X}}|L_{2500}, z)\, N_{\text{tot}}\, \Delta {\log_{10}}L_{\text{X}}\\
                     &= \frac{N_{\text{det}}\,N_{\text{tot}}}{\sum_{i=1}^{N_{\text{tot}}}p({\text{det}}|L_{{\text{X}}}, z_i)}.
    \end{split}
\end{equation}
In the limit where $p({\text{det}}|L_{{\text{X}}}, z)=1$ for all quasars in a given {\lx} bin (i.e. the X-ray data are sufficiently deep that any quasar with that {\lx} should be detected), $N_{\text{corr}} = N_{\text{det}}$ and thus corresponds to the "uncorrected" (solid) histograms in Figure~\ref{fig:binnedLx}. Thus, as expected, at the highest {\lx} the corrected (open histograms/error bars) and uncorrected (solid histograms) estimates are consistent.

The corrected counts are plotted in Fig.~\ref{fig:binnedLx} as coloured outlined histograms and Poisson errors are generated based on the $N_{\text{det}}$ in each bin and applying Gehrels' method for small number statistics \citep{gehrels_confidence_1986}. As expected, the correction is larger at low X-ray luminosities. The corrected {\lx} distribution is perhaps Gaussian with the centre, $\mu$, and width, $\sigma$ potentially varying with {\luv} and $z$; however, this model can only correct bins with $N_{\text{det}}>0$ and significant binning is required. Additionally, while narrower bins leads to higher resolution, the uncertainties on {\lx} are comparable to the {\lx} bin-width. In the following section we move on to using Maximum Likelihood Estimation (MLE) to arrive at a fully unbinned approach to determining the {\lx} distribution as a function of {\luv} and $z$.

\subsection{Maximum likelihood fitting}
\label{sec:mle}
The observed and corrected distributions in Fig.~\ref{fig:binnedLx} suggest that $\log_{10}${\lx} is normally distributed for quasars of a given {\luv} and $z$:
\begin{equation}
    P(L_{\text{X}} | L_{2500}, z) = \frac{1}{\sigma\sqrt{2\pi}} \, 
    {\text{exp}}\left[-\frac{({\log_{10}}L_{\text{X}} - \mu)^2}{2\sigma^2}\right],
\label{eq:Plx}
\end{equation}
with mean $\mu$ and width $\sigma$ both of which may depend on {\luv} and/or $z$. In this section, we will attempt to fit the X-ray luminosity distribution function from equation~\ref{eq:Plx} via maximum likelihood estimation (MLE) and will investigate the requirement for {\luv}- and $z$-dependence. 

The log-likelihood (which we derive in Appendix~\ref{app:loglike}) is given by
\begin{equation}
    \begin{split}
        \ln \mathcal{L}(\theta) = &\sum_{i=1}^{N_{\text{det}}}\ln \int_{40}^{\infty} P(L_{{\text{X}}}|L_{2500_i}, z_i, \theta)\ P(N_i|N_{\text{exp}}) \ {\text{d}}{\log_{10}}L_{\text{X}}\ + \\
        &\sum_{j=1}^{N_{\text{not}}}\ln \int_{40}^{\infty} P(L_{{\text{X}}}|L_{2500_j}, z_j, \theta)\ p(\overline{\text{det}}|L_{\text{X}}, z_j)\ {\text{d}}{\log_{10}}L_{\text{X}},
    \end{split}
    \label{eq:lnlike}
\end{equation}
where the first and second terms account for the X-ray detected and undetected quasar samples, respectively. Considering the X-ray detected term, $P(L_{{\text{X}}}|L_{2500_i}, z_i, \theta)$ is the probability of detected quasar $i$ having an X-ray luminosity {\lx} (drawn from the corresponding log-normal distribution) given its UV luminosity $L_{2500_i}$ and redshift $z_i$ calculated from equation~\ref{eq:Plx} with parameters $\theta=\mu,\sigma$. The $P(N_i|N_{\text{exp}})$ term takes into account the uncertainty on the measured X-ray luminosity of the quasar, which is described by a Poisson distribution:
\begin{equation}
    P(N_i|N_{\text{exp}}) = \frac{N_{{\text{exp}}}^{N_i}}{N_i!} \, e^{-N_{{\text{exp}}}}
    \label{eq:Pni}
\end{equation}
with $N_i$, the total observed counts for quasar $i$, and $N_{\text{exp}}$, the expected number of counts from a source with {\lx} which is determined via
\begin{equation}
    N_{\text{exp}} = \frac{L_{{\text{X}}}}{4\pi D_L^2\left(z_i\right) K_{\text{corr}}\left(z_i\right)}\times{\text{ECF}}_i\times{\text{EEF}}\times t_{{\text{exp}}_i} + B_i.
    \label{eq:nexp}
\end{equation}
The energy conversion factor, ECF, exposure, $t_{\text{exp}}$, background counts, $B_i$, and total counts, $N_i$ are specific to each X-ray detection and calculated during the reduction. \hl{The encircled energy fraction, EEF, is 70\,\% for our adopted aperture and is based on the point spread function at 2\,keV, the average energy (weighted by the response) of the full band}. The luminosity distance, $D_L(z_i)$, and K-correction, $K_{\text{corr}}(z_i)$, are calculated for the redshift $z_i$ of the quasar. Since the X-ray luminosities (fluxes more precisely) and errors are calculated from a Poisson distribution by {\sc xmmpype}, $P(N_i|N_{\text{exp}})$ will be maximal when the integration variable {\lx} equals the estimated X-ray luminosity of quasar $i$, $L_{{\text{X}}_i}$.
We note that the maximum of $P(N_i|N_{\text{exp}})$ corresponds to our nominal best estimate of the X-ray luminosity, $L_{\text{X}_i}$, for a given detected quasar.

The X-ray undetected term, similarly to the detected term, depends on the probability of undetected quasar $j$ having an X-ray luminosity {\lx} given its UV luminosity $L_{2500_j}$ and redshift $z_j$, $P(L_{{\text{X}}}|L_{2500_j}, z_j, \theta)$. This probability is multiplied by the probability of quasar $j$ remaining undetected if it were to have X-ray luminosity {\lx}:
\begin{equation}
    p(\overline{\text{det}}|L_{\text{X}}, z_j) = 1 - p({\text{det}}|L_{\text{X}}, z_j)
    \label{eq:pndet}
\end{equation}
where $p({\text{det}}|L_{\text{X}}, z_j)$ is calculated from the area curves (Fig.~\ref{fig:acurve}) via
\begin{equation}
    p({\text{det}}|L_{\text{X}}, z_j) = \frac{{\text{Area}}}{{\text{Total\ Area}}}.
    \label{eq:pdet}
\end{equation}

The integration limits are set as ${\log_{10}}L_{\text{X}}=40,\infty$. In practise the upper limit is set by the maximum X-ray flux probed by the sensitivity curves ($10^{-10}$\,erg\,s$^{-1}$\,cm$^{-2}$). 

\subsection[Distribution of \texorpdfstring{\lx}{Lx} in fixed \texorpdfstring{\luv}{L2500} and redshift bins]{Distribution of $\vect{L_{\text{X}}}$ in fixed $\vect{L_{2500}}$ and redshift bins}
\label{sec:luvzbin}
\begin{figure}
    \centering
    \includegraphics[width=\linewidth]{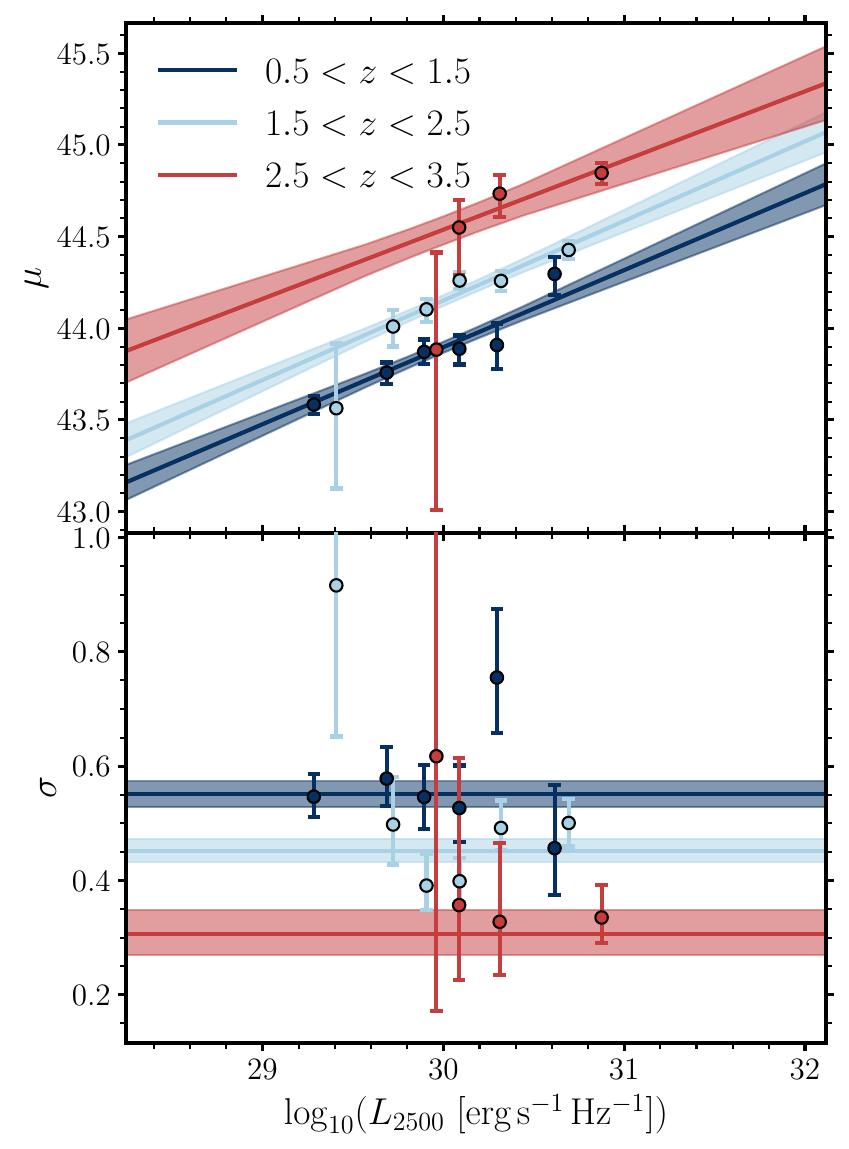}
        \caption{Model parameters $\mu$ (top) and $\sigma$ (bottom) from equation~\ref{eq:Plx} and estimated via MLE as a function of {\luv}. The parameters for model (i), which requires running the MLE on data binned by {\luv} and $z$, are shown by the circles and 1-$\sigma$ error bars. Colours represent $z$ bins. Not all $(z, L_{2500})$ bins contain data and so some bins are missing from the analysis. The lines and shaded regions correspond to the $\mu$ and $\sigma$ parameters obtained when modelling a linear dependence of {\luv} on $\mu$ and no dependence for $\sigma$ (model (ii)). Significant trends with {\luv} exist for $\mu$ in all redshift bins and the results from models (i) and (ii) match well. Within the errors, models (i) and (ii) agree with a {\luv}-independent $\sigma$ but a clear decreasing $\sigma$ with increasing redshift.}
    \label{fig:mu_sig_3}
\end{figure}

We first aim to determine if and how the X-ray luminosity distribution changes as a function of {\luv} and $z$. We divide X-ray detected and undetected quasar samples between equally spaced redshift bins: $0.5 < z < 1.5$, $1.5 < z < 2.5$, and $2.5 < z < 3.5$. We also split the samples across six {\luv} bins such that there are approximately equal numbers of quasars in each bin.
For each $(z_k, L_{2500_l})$ bin we fit for $\mu$ and $\sigma$ by maximising the log-likelihood in equation~\ref{eq:lnlike} with the Python package {\sc emcee} \citep{emcee2013}. The best-fitting parameters are presented in Fig.~\ref{fig:mu_sig_3} as circles. $\mu$ is clearly dependent on {\luv} with the mean {\lx} increasing with increasing {\luv} across all redshift bins. On the other hand, there is little evidence for a {\luv}-dependent $\sigma$ at any redshift but it is possible that $\sigma$ decreases as redshift increases suggesting that the distribution of {\lx} is narrower at greater redshifts. In light of these correlations, we remove the {\luv} binning in the next section and model the relationship between {\luv} and $\mu$ (and $\sigma$) as linear. 

\begin{table*}
    \centering
    \caption{The models fitted with MLE. Columns are, in order, model number for referring to in text; the binning required of the model and whether or not there is redshift evolution for the completely unbinned models; the parameters of the model; the number of dimensions of the model which takes into account the number of redshift and {\luv} bins; the AIC values; $\Delta$AIC is the difference between the AIC for that model and the lowest AIC value.  \hl{$\log_{10} Z$ is the PyMultiNest Bayesian evidence for each of the unbinned models normalised by the most likely model (highest evidence).} Model (vii) with the lowest AIC is presented as bold.}
    \begin{tabular}{c|c|c|c|c|c|c}
\hline
Model & Binning & Parameters & $N_{\text{dim}}$ & AIC & $\Delta$AIC & \hl{$\log_{10} Z$}\\
\hline
 (i) & $z$, {\luv} & $(\mu$, $\sigma)$ for each {\luv} and $z$ bin & 36 & 10131.86 & 99.20 & \\
 (ii) & $z$ & $(m_{\mu}$, $c_{\mu}$, $m_{\sigma}$, $c_{\sigma})$ for each $z$ bin & 12 & 10090.76 & 58.10 & \\
 (iii) & $z$ [constant $\sigma$ with {\luv}] & $(m_{\mu}$, $c_{\mu}$, $\sigma)$ for each $z$ bin & 9 & 10085.77 & 53.12 & \\
 (iv) & Unbinned, no $z$ evolution & $m_\mu$, $c_\mu$, $\sigma$ & 3 & 10145.90 & 113.25 & -22.3\\
 (v) & Unbinned, no $z$ evolution & $m_\mu$, $c_\mu$, $m_\sigma$, $c_\sigma$ & 4 & 10145.40 & 112.74 & -24.2\\
 (vi) & Unbinned, $z$ evolution & $m_\mu(z)$, $c_\mu(z)$, $\sigma$ & 5 & 10061.18 & 28.53 & -5.8\\
 \textbf{(vii)} & \textbf{Unbinned, \bm{$z$} evolution} & \textbf{\bm{$m_\mu$}, \bm{$c_\mu(z)$}, \bm{$\sigma(z)$}} & \textbf{5} & \textbf{10032.65} & \textbf{0.00} & \textbf{0.0}\\
 (viii) & Unbinned, $z$ evolution & $m_\mu(z)$, $c_\mu(z)$, $\sigma(z)$ & 6 & 10032.67 & 0.01 & -2.4\\
\hline

    \end{tabular}
    \label{tab:AIC}
\end{table*}

\subsection[\texorpdfstring{\luv}{L2500}-dependent distribution of \texorpdfstring{\lx}{Lx} in fixed redshift bins]{$\vect{L_{2500}}$-dependent distribution of $\vect{L_{\text{X}}}$ in fixed redshift bins}
\label{sec:zbins}
When binning by {\luv} the model parameter $\mu$ (i.e. the average of the $\log$ \lx\ distribution) appears to increase as {\luv} increases. We model this dependence on {\luv} for $\mu$ and $\sigma$ as linear with $\log_{10}${\luv}:
\begin{equation}
    \begin{split}
        \mu &= m_\mu ({\log_{10}}L_{\text{2500}} - 30) + c_\mu; \\
        \sigma &= m_\sigma ({\log_{10}}L_{\text{2500}} - 30) + c_\sigma.
    \end{split}
\end{equation}
We perform MLE on the $z$-binned data to constrain $m_\mu$, $m_\sigma$, $c_\mu$, and $c_\sigma$ (model (ii)), thus removing the need to bin our quasar sample according to {\luv}. It is not clear that $\sigma$ varies with {\luv}, thus we repeat the MLE for the model where $\sigma$ does not depend on {\luv}, formally $m_\sigma=0$ therefore $\sigma=c_\sigma$ (model (iii)). 

To compare the different models (with different numbers of free parameters) we will use the Akaike Information Criterion defined as
\begin{equation}
    \text{AIC} = 2N_{\text{dim}} - 2\ln\hat{\mathcal{L}}
\end{equation}
with $N_{\text{dim}}$ the number of free parameters and $\hat{\mathcal{L}}$ the maximum of the likelihood function (Equation~\ref{eq:lnlike}). The AIC penalises models with a large number of parameters and models with lower AICs are considered to better represent the data. To calculate the AICs of the models with binning, we treat the model as a piecewise function such that the maximum log-likelihood is the sum of the maximum log-likelihood over all $z$ bins (and {\luv} bins for model (i)) and $N_{\text{dim}}$ is the total number of parameters across all bins.
Model (iii) is formally a better fit with a lower AIC than model (ii) (see Table~\ref{tab:AIC}) and so we plot the results of model (iii) in Fig.~\ref{fig:mu_sig_3} as the straight lines and shaded regions. Within the errors the linear dependence on {\luv} agrees with the {\luv}-binned fits of model (i) (circles; Section~\ref{sec:luvzbin}).

Across redshift bins, the intercept of the $\mu$ relation and $\sigma$ in general change. At a given {\luv}, $\mu$ increases as redshift increases and $\sigma$ decreases. This can be seen more clearly in Fig.~\ref{fig:zevo_param}. The gradient of the $\mu$-{\luv} relation appears to be relatively constant with redshift; however, $c_\mu$ is clearly increasing as redshift increases and $\sigma$ is decreasing. We thus move on to model the dependence of these parameters on redshift to arrive at a fully unbinned MLE in Section~\ref{sec:zdep}.

\subsection{Continuous model of the redshift evolution}
\label{sec:zdep}
In this section we arrive at a selection of models to describe the whole data sample in a continuous manner instead of discrete {\luv} or $z$ bins. We model any possible redshift evolution of $\mu$ and $\sigma$ via a linear dependence of $m_\mu$, $c_\mu$, and $\sigma$ on $z$. We perform the MLE with various models with different $z$ dependencies, explicitly:
\begin{enumerate}
    \setcounter{enumi}{3}
    \item no redshift evolution, with parameters $m_\mu$, $c_\mu$, and $\sigma$. This is the equivalent of model (iii) in the limit of one redshift bin.
    \item no redshift evolution, with parameters $m_\mu$, $c_\mu$, $m_\sigma$, and $c_\sigma$. This is the equivalent of model (ii) but assuming a single, broad redshift bin.
    \item only redshift evolution of $\mu$, with gradient and intercept parameters for $m_\mu(z)$ and $c_\mu(z)$, and a constant $\sigma$.
    \item redshift evolution of only $c_\mu$ and $\sigma$ with gradient and intercept parameters for $c_\mu(z)$ and $\sigma(z)$ and a constant $m_\mu$.
    \item redshift evolution of $m_\mu$, $c_\mu$, and $\sigma$ with gradient and intercept parameters for $m_\mu(z)$, $c_\mu(z)$, and $\sigma(z)$.
\end{enumerate}

\begin{table}
    \centering
    \caption{Best fit parameter values and 1-$\sigma$ uncertainties for model (vii) where $\mu(L_{2500},z) = m_\mu(\log_{10}L_{2500} - 30) + p_{\mu}z + k_{\mu}$ and $\sigma(z)=p_{\sigma}z + k_{\sigma}.$}
    \begin{tabular}{c|c}
\hline
Parameter & Value \\
\hline
$m_\mu$ & $0.313_{-0.034}^{+0.035}$ \\
$p_{\mu}$ & $0.414_{-0.033}^{+0.033}$ \\
$k_{\mu}$ & $43.426_{-0.057}^{+0.056}$ \\
$p_{\sigma}$ & $-0.122_{-0.021}^{+0.022}$ \\
$k_{\sigma}$ & $0.657_{-0.039}^{+0.040}$ \\
\hline
    \end{tabular}
    \label{tab:val}
\end{table}

\begin{figure}
    \centering
    \includegraphics[width=\linewidth]{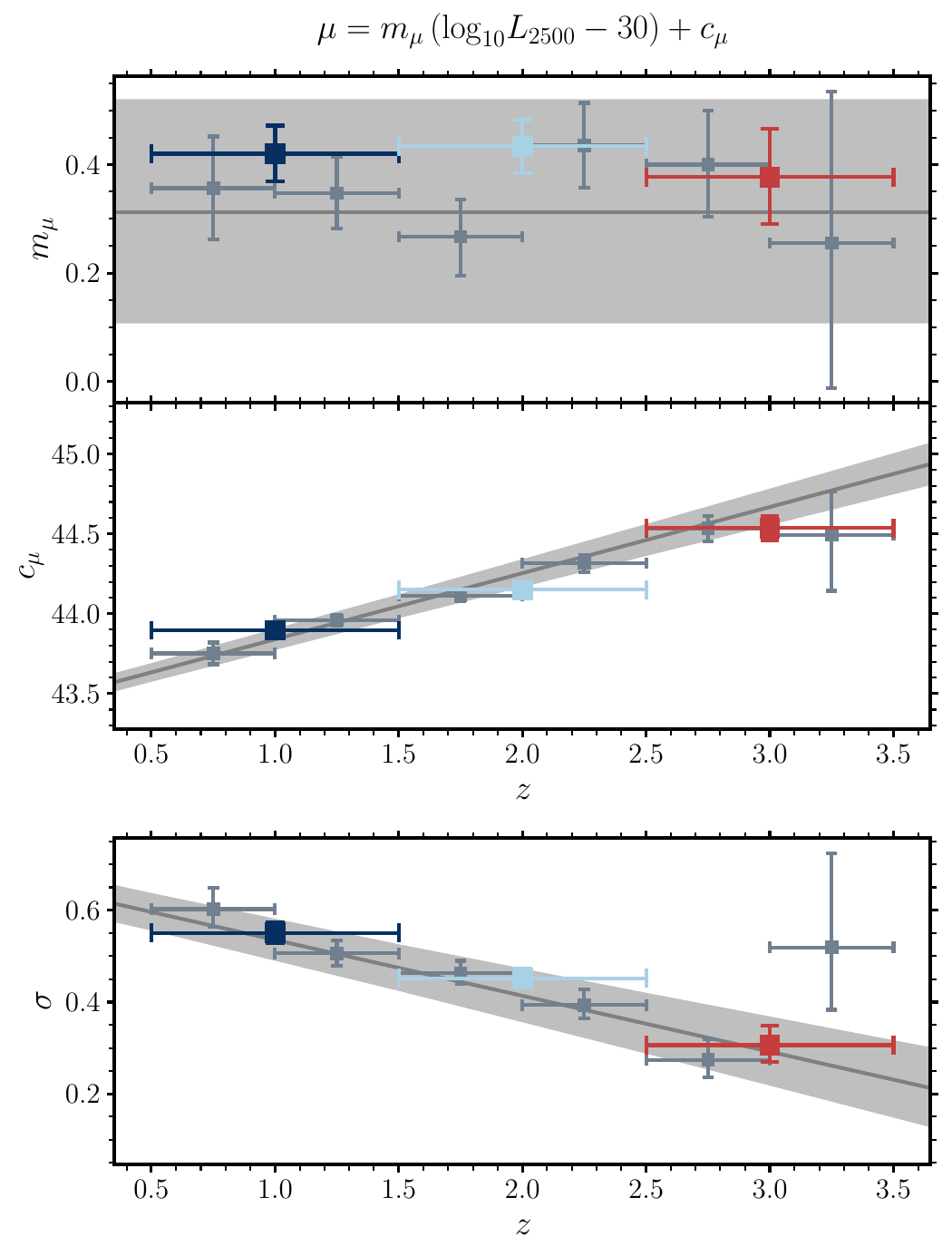}
    \caption{Breakdown of the parameters $\mu$ (top and middle) and $\sigma$ (bottom) as a function of redshift for the $z$-dependent models. The top two panels are the gradient $m_\mu$ and the value of $\mu$ at $\log_{10}(L_{2500})=30$. The squares are the parameter values used to produce the straight-lines in Fig.~\ref{fig:mu_sig_3} from model (ii). The grey lines and shaded regions are the parameter values obtained with model (vii) where redshift evolution is modelled as a linear dependence of $z$ on $m_\mu$, $c_\mu$, and $\sigma$. \hl{The grey squares represent the higher resolution redshift binning used to check the cause of the systematically higher binned points in $m_\mu$.}}
    \label{fig:zevo_param}
\end{figure}

From the AIC values in Table~\ref{tab:AIC}, the model which best represents the data is model (vii) which allows for redshift evolution of $c_\mu$ and $\sigma$ parametrized by,
\begin{equation}
    \begin{split}
        c_\mu &= p_{\mu} z + k_{\mu};\\
        \sigma &= p_{\sigma} z + k_{\sigma}.
    \end{split}
\end{equation}
\hl{We also make use of nested sampling via the {\sc MultiNest} algorithm \citep{feroz_2008_multimodal, feroz_2009_multinest, feroz_2019_importance} and its Python implementation {\sc PyMultiNest} \citep{buchner_2014_pymultinest} for model comparison assessed via the Bayesian evidence which also prefers model (vii) (see the 7th column of Table~\ref{tab:AIC}). In fact, there is greater evidence for the models with redshift evolution (vi--viii) compared to those without (iv, v).}
The grey lines and shaded regions in Fig.~\ref{fig:zevo_param} are the best-fitting parameters for model (vii) (also listed in Table~\ref{tab:val}). The $z$-binned $m_\mu$ values from Section~\ref{sec:zbins} are systematically higher than the continuous redshift modelling in this section. This is due to the distribution of objects within the relatively broad redshift bins and the intrinsic redshift evolution of $\mu$ within such bins. The higher {\luv} and {\lx} sources within a given redshift bin are preferentially identified toward higher redshifts and thus in our binned results a steeper relation between {\lx} and {\luv} (i.e. a steeper $m_\mu$) is recovered to account for this redshift evolution. The intercept, $c_\mu$, does not have such a strong dependence on the width of the redshift bin. Increasing the number of redshift bins by a factor of two removes this systematic bias but also reduces the number of objects in each bin and thus leads to greater statistical uncertainties in the parameters.

\section[Underlying \texorpdfstring{\luv--\lx}{L2500-Lx} distribution]{Underlying $\vect{L_{2500}}$--$\vect{L_{\text{X}}}$ distribution}
\label{sec:LuvLx}

With model (vii) in hand, for a given $z$, as expected the peak of the intrinsic {\lx} distribution increases as {\luv} increases. Perhaps not as obvious is that for a given {\luv}, the intrinsic {\lx} distribution shifts to higher {\lx} as redshift increases ($c_\mu$ increases since the gradient of $c_\mu(z)$ is found to be positive) and also narrows ($\sigma$ decreases since the gradient of $\sigma(z)$ is negative). 

We compare the underlying distribution of {\lx} of our optically-selected quasar sample to the observed data and original binned corrections in Fig.~\ref{fig:binnedLx}. For each object in our sample with a given {\luv} and redshift, detected or otherwise, we draw 100 samples from the distribution function (equation~\ref{eq:Plx}) with the best-fit parameters listed in Table~\ref{tab:val}, effectively creating a mock sample of {\lx} measurements if there were no limitations in X-ray depth. In Fig.~\ref{fig:binnedLx} we then normalise to the number of objects in each {\luv} and redshift bin (black histograms). Unlike the binned corrections, we can infer the source counts in {\lx} bins with zero observed sources. In the majority of {\luv} and $z$ bins the binned corrections and the MLE corrections agree. However, in some panels (e.g., second row, second to last column) the binned corrections are significantly lower than the MLE distribution which we believe to be the combination of using the Bayesian sensitivity curves which appropriately account for Eddington bias but are not suitable for the crude binned corrections carried out in Section~\ref{sec:binned} in bins where the majority of the sample is around the flux limit. As noted previously, in all bins, the detected sources are only probing the high {\lx} tail of the distribution.

\begin{figure*}
    \centering
    \includegraphics[width=\linewidth]{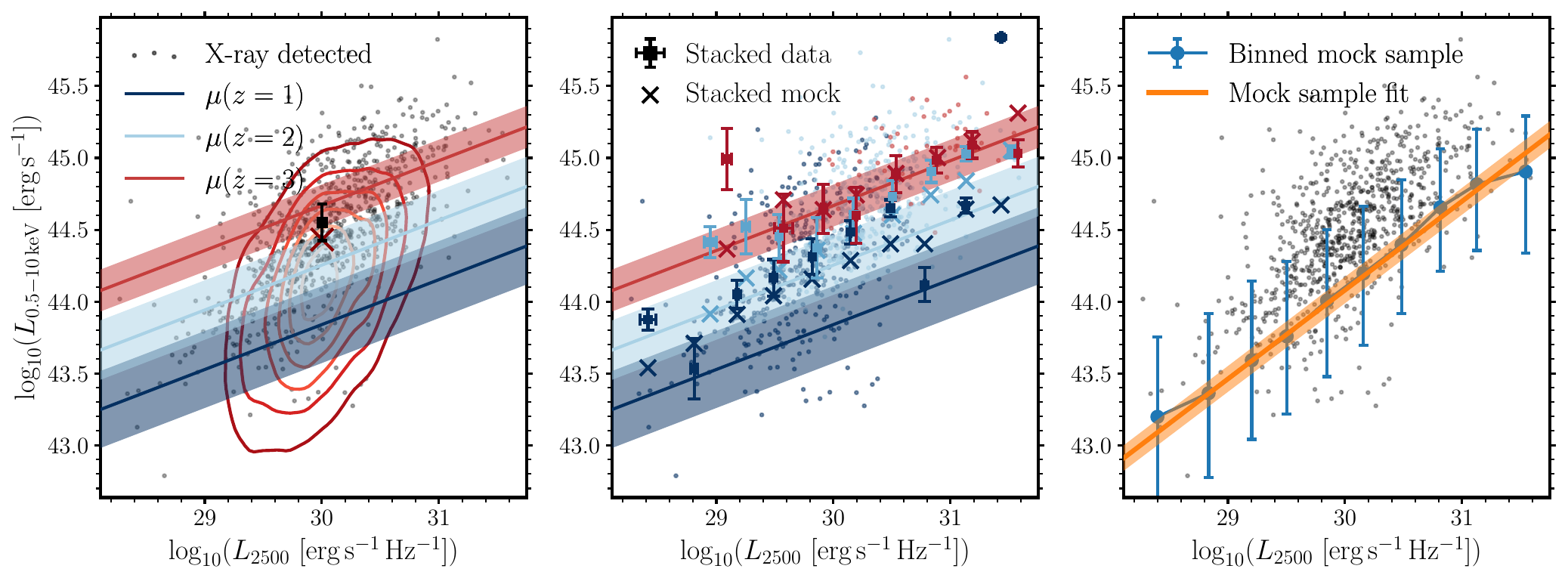}
    \caption{The {\luv}--{\lx} plane with the X-ray detected sample plotted as black circles. Left: The solid blue and red lines are the peaks of the {\lx} distribution function with constant $z=1,2,3$ and the shaded areas are the 0.5-$\sigma$ width of the distribution.
    The mock sample is represented by the red contours. The black square is the result from our XMM stacking analysis and red cross is the equivalent stacking of the mock sample. Middle: The same relations and data as in the left panel, now overplotted with the stacked data (squares) and stacked mock sample (crosses) in redshift and {\luv} bins. Right: The orange line is produced by fitting a straight line to the mock sample, and the blue points and error bars are the median {\lx} in {\luv} bins and the 1-$\sigma$ of the {\lx} distributions, respectively.}
    \label{fig:LuvLx}
\end{figure*}

As mentioned in Section~\ref{sec:intro}, there is a well-known correlation between the optical and X-ray luminosities of AGN with $L_{\text{X}}\propto L_{\text{UV}}^{\gamma}$ and $\gamma\sim0.6$. In this work, we have found that the X-ray luminosity distribution is a function of {\luv} and redshift, with the peak of the distribution given by $\mu(L_{2500},z)$. In the left panel of Fig.~\ref{fig:LuvLx} we plot the peak of the {\lx} distribution for a constant $z=1,2,3$ where an increase in redshift produces a higher {\lx} for a given {\luv}.

In order to check our results, we produce a stacked value of {\lx} from the X-ray counts extracted at the positions of all of our quasars. We do so by calculating individual X-ray luminosities for each optical source and then produce a mean {\lx}. This produces an {\lx} value (black square in the left panel of Fig.~\ref{fig:LuvLx}) that is higher than the centre of the contours due to the mode of a log-normal distribution (which the {\lx} distribution is) being different from its mean. As a sanity check, calculating the average {\lx} in the same way from the mock data used to produce the contours results in a higher {\lx} value than the contours would suggest (red cross); however, it is consistent with the stacked {\lx} from the data. We do the same in the middle panel of Fig.~\ref{fig:LuvLx} in bins of {\luv} and $z$ to compare to the relations. Again, we find that the {\lx} values from the stacked data are systematically higher than the relations and, although suffering from small-number statistics with this relatively high-resolution binning, the stacked mock data is in agreement. In fact, the stacked data in the highest redshift bin (red squares) appear to agree too well with the relations; however, this is due to the distribution of redshifts within this redshift bin. If instead we were to plot the relations for the mean redshift within each redshift bin, the red line would shift lower and the squares would be offset. The redshift dependence on the width of the {\lx} distribution is also having an effect: at low redshifts where the distribution is wider, the discrepancy between the stacked data and the relation is greater. This in turn reduces the redshift dependence in the stacked data.

In order to compare more directly to the literature, we take the mock sample (red contours in the left-hand panel of Fig.~\ref{fig:LuvLx}) and fit a straight line and we obtain a $\gamma\simeq0.62$ which is in agreement with the literature (right-hand panel of Fig.~\ref{fig:LuvLx}). It is not obvious that this best-fit line is in agreement with the contours; however, the median $\log_{10}${\lx} in {\luv} bins is consistent with the $\gamma\simeq0.62$ relation (blue points and error bars in the right panel of Fig.~\ref{fig:LuvLx}).

\section[Corrected \texorpdfstring{\aox}{aox}]{Corrected $\vect{\alpha_{\text{ox}}}$}
\label{sec:aox}
\begin{figure}
    \centering
    \includegraphics[width=\linewidth]{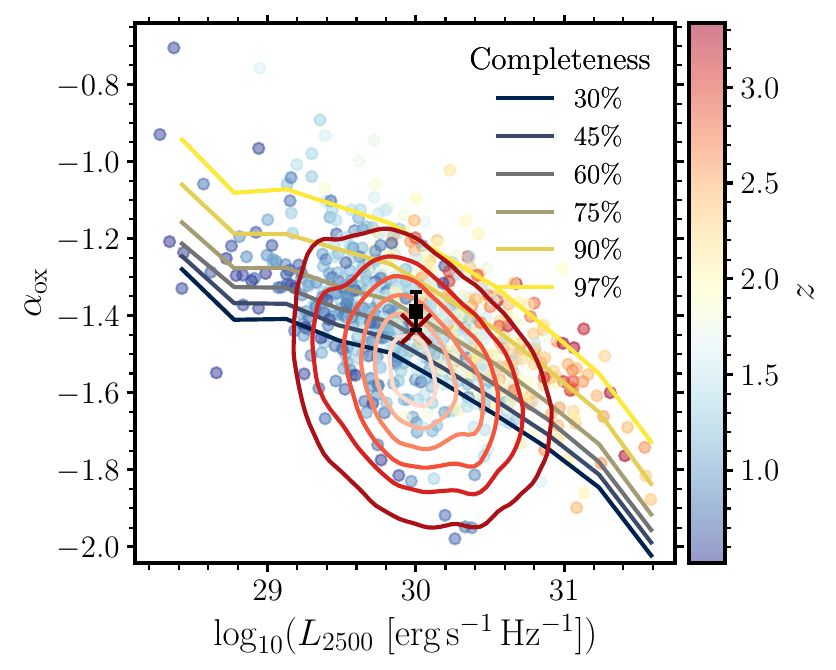}
    \caption{{\aox} versus {\luv} for our X-ray detected quasar sample with points colour-coded by redshift. The blue-to-yellow lines correspond to the completeness of our sample across {\luv}. The {\luv} distribution is plotted above the axes. The red contours are the mock sample. The black square and error bars is the {\aox} result from our XMM stacking analysis. The red cross is generated by stacking the random sample from the MLE results used to produce the contours.}
    \label{fig:aox}
\end{figure}

The spectral slope between the X-ray and optical, {\aox}, is often used as a means of describing the relationship between the X-ray and UV luminosities, and is calculated as follows,
\begin{equation}
    \alpha_{\text{ox}} = \frac{\log_{10}\left(L_{2{\text{keV}}} / L_{2500{\textup{\AA}}}\right)}{\log_{10}\left(\nu_{2{\text{keV}}} / \nu_{2500{\textup{\AA}}}\right)}
    \label{eq:aox}
\end{equation}
where $L_{2{\rm\,keV}}$ is the monochromatic X-ray luminosity at 2\,keV and corresponding frequency $\nu_{2\text{keV}}$. $\nu_{2500\textup{\AA}}$ is the frequency equivalent to 2500\,\AA. We calculate {\aox} for our X-ray detected sample by converting the full band $L_{0.5-10{\rm\,keV}}$ into $L_{2{\rm\,keV}}$ and plot these values against {\luv} in Fig.~\ref{fig:aox} colour-coded by redshift. The flux-limited nature of the parent sample is clear here in that the highest {\luv} quasars are only found at high redshifts; however, the completeness curves generated from the sensitivity curves in Fig.~\ref{fig:acurve} reveal that the completeness of our optically-selected sample drops significantly as X-ray luminosity decreases ({\aox} decreases) across the full range of {\luv} probed by our quasar sample. 

In what follows, we make use of the derived underlying X-ray luminosity distribution as a function of both {\luv} and redshift to produce a corrected {\luv}--{\aox} relation. We calculate {\aox} for the mock sample in Section~\ref{sec:LuvLx} with equation~\ref{eq:aox} and produce the red contours in Fig.~\ref{fig:aox}. The true underlying {\aox} distribution suggests that we are missing the {\luv}-moderate, {\lx}-faint population which any {\aox} relation should account for.

In order to check our results, we produce a stacked value of {\aox}. We take the stacked {\lx} value from Section~\ref{sec:LuvLx} for all of the quasars, log this value and convert to {\aox} with the mean $\log_{10}L_{2500}$ of our data. This produces an {\aox} value (black square in Fig.~\ref{fig:aox}) that is higher than the centre of the contours due to the mode of a log-normal distribution (which {\aox} is) being different from its mean. Calculating the average {\aox} in the same way from the mock sample used to produce the contours results in a higher {\aox} value than the contours would suggest; however, it is consistent with the stacked {\aox} from the data. We caution that simple linear stacked measurements to infer relations with broad, log-normal shapes will not correspond to the peak (mode) of the distribution but be biased high as we have found here.

The relationship between the peak of the {\aox} distribution and the {\luv} and redshift is given by
\begin{equation}
    \alpha_{\text{ox}}(L_{2500}, z) = a \log_{10}\left(\frac{L_{2500}}{{\text{erg}}\,{\text{s}}^{-1}\,{\text{Hz}}^{-1}}\right) + b z + c,
    \label{eq:aoxr}
\end{equation}
where $a=-0.264^{+0.013}_{-0.013}$, $b=0.159^{+0.013}_{-0.013}$ and $c=6.095^{+0.400}_{-0.395}$. In short, the relation is derived by converting the peak of the full-band {\lx} (0.5--10\,keV), given by model (vii) with the parameter values from Table~\ref{tab:val}, to the 2\,keV monochromatic luminosity and substituting this in equation~\ref{eq:aox}. The full derivation is presented in Appendix~\ref{app:aoxr}. Thus far we have not considered whether the parameters of our model are independent; however, 
parameters $a$ and $c$ are correlated since both are functions of $m_\mu$ and so the uncertainty on {\aox} will include, at the very least, the covariance of $a$ and $c$. We consign the equation for $\Delta\alpha_{\text{ox}}$ and its derivation to Appendix~\ref{app:aoxr} but note here that we assume the parameters of our model in Table~\ref{tab:val} are independent (but see Appendix~\ref{app:aoxr} and Fig.~\ref{fig:corner}). We provide the posterior distributions of the parameters as supplementary data.

\begin{figure*}
    \centering
    \includegraphics[width=\linewidth]{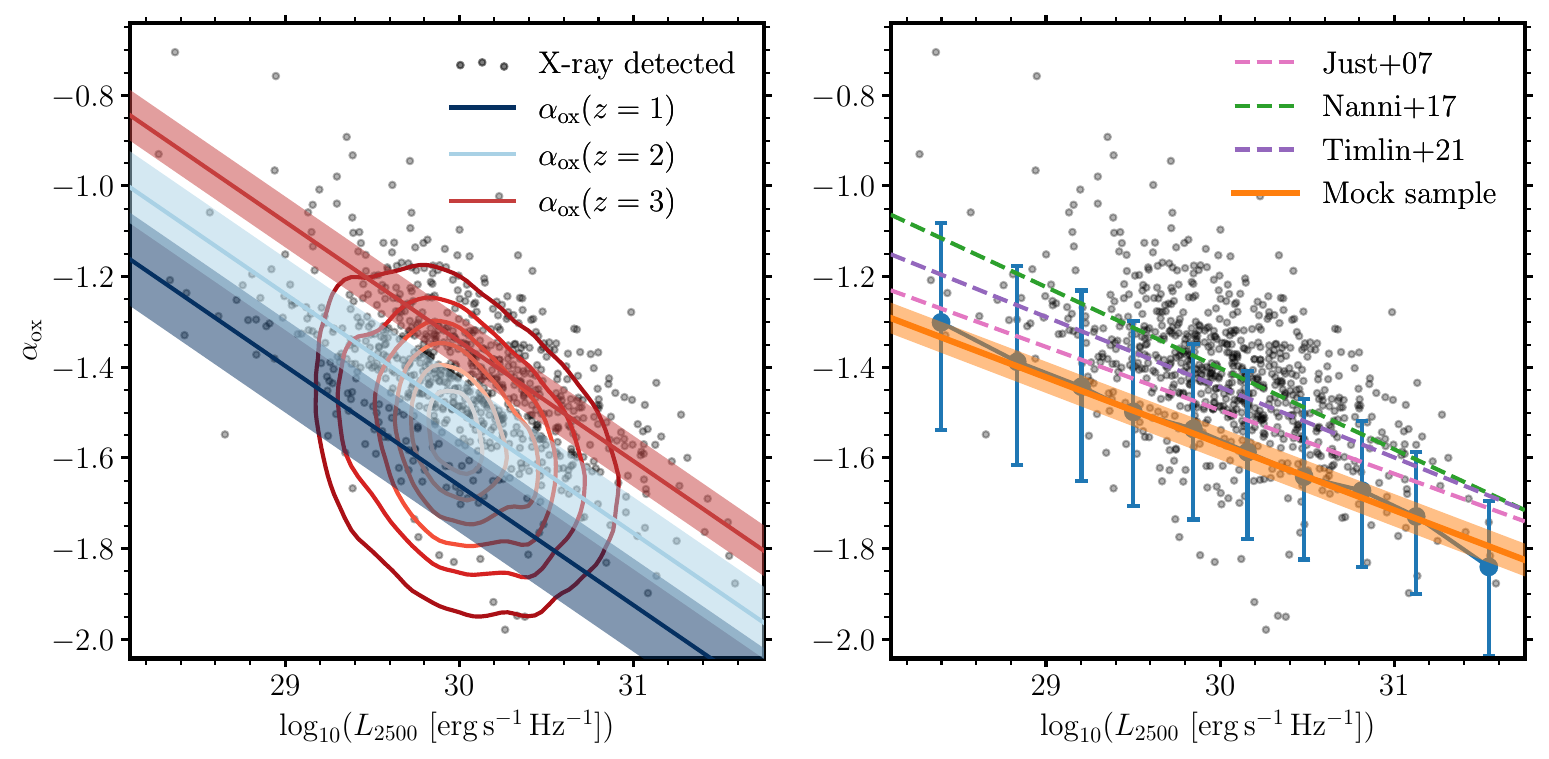}
    \caption{The same {\luv}--{\aox} space as Fig.~\ref{fig:aox} with the X-ray detected sample as black circles and mock sample described by the contours. Left: The solid blue and red lines are calculated via equation~\ref{eq:aoxr} with constant $z=1,2,3$ and the shaded areas are the 0.5-$\sigma$ widths of the relation.
    Right: The dashed lines are relations from the literature. The orange line is produced by fitting a straight line to the mock sample which is consistent with the median {\aox} in {\luv} bins (blue points and error bars).}
    \label{fig:aoxr}
\end{figure*}

In Fig.~\ref{fig:aoxr} we plot {\aox} from equation~\ref{eq:aoxr} for a constant $z=1,2,3$ where an increase in redshift produces a vertical shift towards less-negative {\aox} values (upwards on the plot). Our model that describes how the peak of the intrinsic distribution of {\aox} depends on {\luv} at different redshifts (accounting for X-ray sensitivity limits and the underlying redshift evolution of the relation between {\luv} and {\lx} over this redshift range) produces significantly steeper relations (solid lines in Fig.~\ref{fig:aoxr}) than most prior estimates that use X-ray upper-limits and often combine a wide redshift range \citep[e.g., ][as shown by the dashed lines in Fig~\ref{fig:aoxr}, right]{just_xray_2007, nanni_x-ray_2017, timliniii_what_2021}. For comparison to the literature, we use our mock sample that corrects for the X-ray incompleteness (but not the uneven sampling of the quasar samples in terms of {\luv} and $z$) and  fit a linear relation between {\luv} and all of our mock {\aox} values. Fitting the mock sample with a single linear relation produces an {\aox} with a flatter slope that is in better agreement with the literature but has a lower normalisation. The black line is lower in normalisation for two reasons: i) it accounts for X-ray fainter sources that tend to be below the sensitivity limits, and ii) it tracks the quasar sample that is dominated by lower redshift ($z\lesssim2$) sources, which we find to have lower {\aox} (at a given {\luv}).

\hl{We consider the effect of a soft excess which is thought to be important at restframe energies below 1\,keV \citep{halpern_1984_variable, arnaud_1985_exosat, done_2012_intrinsic} Thus it can only be observed in our lowest redshift sources (0.75\,keV at $z=0.5$). None the less, we conducted simulations with \textsc{xspec} \citep{arnaud_1996_xspec}, finding that at $z=0.5$ we could be overestimating the 2\,keV luminosities by only 10\,\% which is insignificant when compared to the redshift dependence suggested by our results. Additionally, we see no evidence for a soft excess in the hardness ratios of our sample (Fig.~\ref{fig:HR}).}

\hl{Although we are aiming to produce an {\luv}--{\aox} relation corrected only for observational selection effects and are not considering any intrinsic absorption, we investigate if the observed redshift-dependence can be explained by intrinsic absorption of the X-ray emission. We draw intrinsic $\log n_\text{H}$ values uniformly between 20 and 22, and $\Gamma$ from a normal distribution centred on 1.9 with a standard deviation of 0.2. We produce X-ray spectra with these parameters using xspec and calculate the observed 2\,keV luminosity from the observed 0.5--10\,keV luminosity. At $z=0.5$, we could be underestimating the 2\,keV luminosity by 20\,\%, 8\,\% at $z=2$, and 5\,\% at $z=3.5$. Although there is a systematic underestimation that is correlated with redshift, the discrepancies are again insignificant compared to the redshift-dependence we observe. Additionally, re-running the maximum likelihood analysis with the sample of \citet{peca_cosmic_2023}, who performed spectral analysis of the Stripe 82 sample to account for intrinsic absorption in their calculations of X-ray luminosities, produces still a redshift-dependent relation.\footnote{\hl{While their source extraction method differs from ours resulting in a different and smaller sample this comparison provides at least a first order test of the effect of intrinsic absorption.}} Hard-band X-ray emission will be less affected by absorption than the full band and so we re-run our whole analysis (including {\sc Nway} matching) on the X-ray sources detected in the hard band and find that model (vii) is still the most successful model in explaining the data. Models (vi) and (viii) are also given a viable joint-second place, but importantly, both of these models involve redshift-dependence. The source numbers are smaller in the hard band, thus we choose not to use this for our main analysis. See Appendix~\ref{app:hard} for the summary statistics using the hard band.} \hl{The SDSS spectra of our quasars also show little evidence of any intrinsic extinction; thus any changes in intrinsic reddening across our redshift range are negligible and cannot explain the observed redshift evolution of the {\luv}--{\lx} relation.}

\section{Discussion}
\label{sec:disc}

Using a sophisticated Bayesian framework, we have shown 
that the intrinsic distribution of X-ray luminosities of the SDSS quasar sample evolves with redshift, shifting toward higher {\lx} at a given {\luv} and with decreasing scatter in the distribution as redshift increases. Our finding is in disagreement with a number of prior works that do not find any evolution in this relation, albeit for distinct samples and without applying the sophisticated analysis techniques that we present here (see Section~\ref{sec:intro}). However, \citet{kelly_evolution_2007} also find evolution of the {\aox}--$z$ relation in a sample of radio-quiet quasars across $z=0.1$--4.7 with {\aox} increasing as redshift increases (with {\aox} depending linearly on the age of the Universe). \hl{Additionally, \citet{shen_soft_2006} also perform a similar maximum likelihood analysis using soft X-ray detection from RASS in the SDSS DR3 and found a redshift dependent {\luv}--{\lx} relation, albeit weaker than found here.\footnote{\hl{The X-ray non-detections were handled differently in their likelihood function (their eq. 16) and $L_{2\,\text{keV}}$ was estimated from the comparatively soft 0.1--2.4\,keV and 0.5-2.0\,keV bands.}}}

\hl{On a quick glance, our redshift-dependent {\luv}--{\lx} relation would suggest that this relation cannot be used as a cosmological probe as \citet{lusso_quasars_2017} suggest, for example. However, we want to stress that our results are applicable only to the overall optically-selected quasar population. Only with carefully chosen sub-samples of quasars can it be possible to use the {\luv}--{\lx} for cosmological purposes \citep{salvestrini_quasars_2019, bisogni_chandra_2021}. Regardless of whether quasars can be used in this way to test cosmological models \citep[see][who suggest the answer is uncertain]{khadka_quasar_2023}, \textit{our} aim is to eventually use this relation to determine the underlying causes of the link between X-ray and UV emission in the larger quasar population.}

As mentioned above, the purpose of this work is to derive the 
intrinsic distribution of {\lx} as a function of {\luv}
and redshift that applies to the optically-selected SDSS quasar sample, specifically.
While our finding of redshift evolution of the {\luv}--{\lx} relation is at odds with the consensus (see Section~\ref{sec:intro}) we do not dwell on complex comparisons as our results apply to a specific (but well-defined) sample. However, one advantage of our work is that we have carefully considered the X-ray sensitivity limitations, without so doing would result in a different answer for the {\luv}--{\lx} relation (or {\aox}). These relations are important for understanding the balance of energetic output coming from the corona versus the accretion disk, and in order to compare to physical models \citep[e.g.,][]{kubota_physical_2018} one must account for X-ray sensitivity limitations of the sample. 

\begin{figure*}
    \centering
    \includegraphics[width=\linewidth]{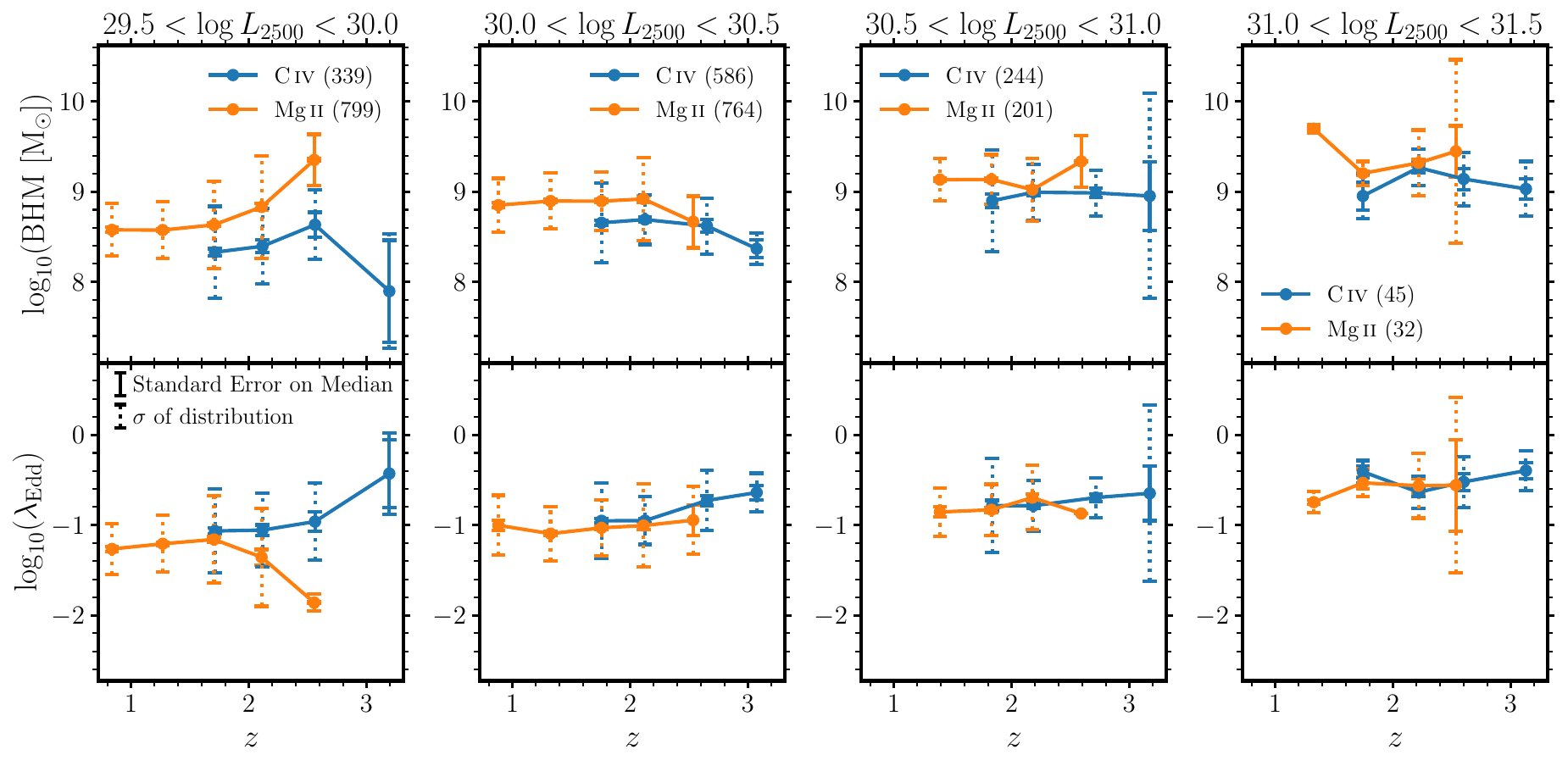}
    \caption{\hl{Black-hole mass (top) and Eddington ratio (bottom) versus redshift for our optically-selected quasar sample in bins of increasing {\luv} (panels left to right). Black hole masses are derived from either the \ion{C}{iv}$\lambda$1550 or \ion{Mg}{ii}$\lambda$2800 emission lines (blue and orange points, respectively) depending on whether the line falls within the spectral coverage at a given redshift. The numbers in the legend refer to the total number of quasars in each luminosity bin with coverage of the \ion{C}{iv} and \ion{Mg}{ii} lines. The error bars mark the standard error on the median (solid) and the standard deviation of the sample in each bin (dotted). We observe no obvious trend in either BHM or $\lambda_\text{Edd}$ with redshift for a given {\luv}.}}
    \label{fig:bhm}
\end{figure*}

While we do not aim to come up with a detailed physical model to explain the observed redshift evolution, it is informative to look at the black hole properties of the optically-selected SDSS quasar sample across redshift and compare to the trends observed in \citet{kubota_physical_2018}. 
We estimate the black hole masses (BHM) and Eddington ratios ($\lambda_\text{Edd}$) of our quasars using the ICA-based spectrum reconstructions (see Fig.~\ref{fig:spec} and Section~\ref{ssec:opt}), calculating BHMs from the full width at half maximum (FWHM) of the \ion{C}{iv}$\lambda$1550 and \ion{Mg}{ii}$\lambda$2800 emission lines, redshift-permitting, and the 1350\,{\AA} and 3000\,{\AA} monochromatic luminosities. For the \ion{C}{iv}-derived BHMs we apply the relation of \citet{coatman_correcting_2017} which accounts for the non-virial component and subsequent asymmetry of the \ion{C}{iv} emission line. The \ion{Mg}{ii} BHMs are estimated with the \citet{vestergaard_mass_2009} relation. We apply bolometric corrections of 5.15 and 3.81 to the monochromatic 1350\,{\AA} and 3000\,{\AA} luminosities, respectively, to estimate the bolometric luminosities \citep{shen_catalog_2011}. With the BHMs and bolometric luminosities in hand, the $\lambda_\text{Edd}$ are calculated. We calculate the median BHM and $\lambda_\text{Edd}$ in bins of redshift along with the standard error on the median values and the standard deviation of the distributions using either the \ion{C}{iv}- and \ion{Mg}{ii}-derived values or both values at redshifts where both lines are within the spectral window  (see Fig.~\ref{fig:bhm}). We do not focus on the absolute values of the quantities but instead focus on the general trends of BHM and $\lambda_\text{Edd}$ with redshift, \hl{observing that neither BHM or $\lambda_\text{Edd}$ show any significant trend with redshift across the majority of the {\luv} bins.}

\hl{The spectral energy distribution (SED) model of \citet{kubota_physical_2018} predicts that the X-ray luminosity scales linearly with BHM, after fixing $L_\text{X}=0.02L_\text{Edd}$ motivated by the SED fits of a handful of AGN \citep[but applied to a larger sample by ][]{mitchell_soux_2023}. In fact, \citet{mitchell_soux_2023} explicitly show that {\aox} (and therefore the {\luv}--{\lx} relation) has a dependence on both the BHM and Eddington-scaled accretion rate due to the relative contributions of the hot X-ray corona, the warm Compton region of the disc and the standard disc in their truncated disc model. Additionally, \citet{kubota_physical_2018} observe that an increase in the Eddington-scaled accretion rate of 1\,dex should produce a \emph{decrease} in $\log${\lx} of 0.5\,dex (for constant {\luv}). The large uncertainties on our BHM and $\lambda_\text{Edd}$ measurements preclude a more detailed discussion on the effect of the BH properties and it is unclear if BH mass and/or Eddington rate are responsible for the observed redshift dependence of the {\luv}--{\lx} relation in our sample. 
In the future, it would be valuable to 
extend the Bayesian hierarchical modelling to consider the underlying Eddington-scaled accretion rate and black hole masses, enabling a more direct comparison with \citet{kubota_physical_2018}.} 

\hl{Quasar SEDs likely evolve with BHM and $\lambda_\text{Edd}$. This true evolution coupled with the flux-limited nature of SDSS leads to quasar samples that have masses and $\lambda_\text{Edd}$ distributions that appear to evolve with redshift. This is implied by the shift to larger BH masses and Eddington ratios as {\luv} increases from left to right in the panels of Fig.~\ref{fig:bhm}. However, these measurements help provide insight into the physical origins of the observed trends. Modelling of the \textit{optical} selection effects will be the focus of a future work. Nevertheless, the lack of significant correlations between the BH properties -- masses and $\lambda_\text{Edd}$ -- and redshift does not appear to be consistent with BHM and/or $\lambda_\text{Edd}$ being responsible for our observed evolution; however, we caveat this with the fact that our measurement uncertainties are large and increase with redshift. Another selection effect in the optical quasar sample or another physical parameter could be responsible for the observed evolution of the relations with redshift.}

\hl{One property of the BHs that we have not considered is their spin. Higher spins have been found to lead to greater X-ray emission relative to the UV emission \citep{temple_testing_2023}. For spin to produce our observed increase in {\lx} as $z$ increases would require the BH spins of our optically-selected sample to increase with redshift. Cosmological simulations suggest that BH spins can evolve through mergers and/or accretion resulting in higher spins at high redshifts \citep[e.g.,][]{dubois_2014_black}; however, \citet{temple_testing_2023} suggest that the spins of SDSS quasars at $z\sim2$ are generally low. Regardless of the cause of the redshift-dependence, a single {\luv}--{\aox} relation with a constant slope across all redshifts is almost certainly not correct and so physical models should not be trying to reproduce a non-evolving {\aox} relation.}

\section{Conclusions}

We have carefully inferred the intrinsic X-ray luminosity distribution as a function of UV luminosity and redshift of the optically-selected SDSS quasars in the Stripe 82 and XXL fields using a sophisticated Bayesian hierarchical modelling approach. We have crossmatched the optical SDSS sample to the XMM point sources with {\sc Nway} \citep{salvato_finding_2018}. We have combined XMM-detected quasars with Bayesian sensitivity curves calculated with the custom {\sc xmmpype} pipeline \citep{georgakakis_serendipitous_2011} in order to extract information from the X-ray undetected quasars. Our main findings are:
\begin{enumerate}
    \item The {\sc xmmpype} reductions produce $\log N$--$\log S$ curves that are consistent with previous works (Fig.~\ref{fig:logNlogS} and Section~\ref{ssec:xray})
    \item The {\luv}--{\lx} relation can be modelled as a Gaussian function with mean $\mu$ which depends on the $\log${\luv} and width $\sigma$ (Section~\ref{sec:mle}).
    \item There is some redshift dependence of the {\luv}--{\lx} relation with $\mu$ increasing with redshift. $\sigma$, on the other hand, decreases as $z$ increases (Section~\ref{sec:zdep} and Fig.~\ref{fig:zevo_param}).
    \item For a constant $z$, our fitted {\luv}--{\lx} relation has a slope of $\gamma\sim0.3$.
    The slope in the observed relation of $\gamma\sim0.6$ found by previous works is reproduced when considering the joint redshift and {\luv} distribution of the optically-selected SDSS quasar sample (Section~\ref{sec:LuvLx} and Fig.~\ref{fig:LuvLx}).
    \item Measurements from stacked X-ray data should be considered with caution when deriving quantities from a log-normal distribution (Sections~\ref{sec:LuvLx} and \ref{sec:aox}).
    \item Attempting to correct the X-ray luminosity distribution in {\lx} bins to account for the undetected quasars can lead to underestimated source counts and is limited to only X-ray luminosity ranges that have been detected (Fig.~\ref{fig:binnedLx} and Section~\ref{sec:binned}). A more sophisticated estimation of the {\lx} distribution is implemented via the Bayesian hierarchical modelling approach used throughout the rest of the paper. 
    \item We produce a relation to describe {\aox} that is now a function of {\luv} and redshift. When marginalising over redshift in our SDSS sample, the {\aox} relation we recover has a slope consistent with the literature but with a lower normalisation (Section~\ref{sec:aox} and Fig.~\ref{fig:aoxr}).
\end{enumerate}

We have made the first steps to understand the intrinsic relationship between the X-ray and UV luminosity by considering the optically-selected SDSS quasar sample. The next step is to approach the problem from an X-ray selected sample in order to parametrize the optical selection. The X-ray selected sample from eROSITA \citep{merloni_erosita_2012, predehl_erosita_2021} with follow-up spectroscopy from SDSS-V \citep{kollmeier_sdss-v_2017}  will be beneficial for this work and support a broader goal of obtaining a full characterisation of the UV and X-ray emission properties of the AGN population and the underlying physical structure of the accreting system that produce them.

\section*{Acknowledgements}
ALR would like to thank Aneesh Naik for helpful discussions surrounding the derivation of the likelihood function, Chris Done for helpful comments on the discussion, and Jack Delaney for help with the $\log N$--$\log S$. We thank Ryan Hickox, Alessandro Peca, and Shiyin Shen for additional comments and the anonymous referee for a comprehensive review.
The research leading to these results has received funding from the
European Union’s Horizon 2020 Programme under the AHEAD2020 project (grant
agreement n. 871158).
ALR and JA acknowledge support from a UKRI Future Leaders Fellowship (grant code: MR/T020989/1). AR acknowledges financial support by the European Union’s Horizon 2020 programme "XMM2ATHENA" under grant agreement No 101004168. AG acknowledges support from the EU H2020-MSCA-ITN-2019 Project 860744 “BiD4BESt: Big Data applications for Black hole Evolution Studies” and the Hellenic Foundation for Research and Innovation (HFRI) project "4MOVE-U" grant agreement 2688.
For the purpose of open access, the authors have applied a Creative Commons Attribution (CC BY) licence to any Author Accepted Manuscript version arising from this submission.

The results in this paper are based on observations obtained with XMM-Newton, an ESA science mission with instruments and contributions directly funded by ESA Member States and NASA.

Funding for the Sloan Digital Sky 
Survey IV has been provided by the 
Alfred P. Sloan Foundation, the U.S. 
Department of Energy Office of 
Science, and the Participating 
Institutions. 

SDSS-IV acknowledges support and 
resources from the Center for High 
Performance Computing  at the 
University of Utah. The SDSS 
website is www.sdss4.org.

SDSS-IV is managed by the 
Astrophysical Research Consortium 
for the Participating Institutions 
of the SDSS Collaboration including 
the Brazilian Participation Group, 
the Carnegie Institution for Science, 
Carnegie Mellon University, Center for 
Astrophysics | Harvard \& 
Smithsonian, the Chilean Participation 
Group, the French Participation Group, 
Instituto de Astrof\'isica de 
Canarias, The Johns Hopkins 
University, Kavli Institute for the 
Physics and Mathematics of the 
Universe (IPMU) / University of 
Tokyo, the Korean Participation Group, 
Lawrence Berkeley National Laboratory, 
Leibniz Institut f\"ur Astrophysik 
Potsdam (AIP),  Max-Planck-Institut 
f\"ur Astronomie (MPIA Heidelberg), 
Max-Planck-Institut f\"ur 
Astrophysik (MPA Garching), 
Max-Planck-Institut f\"ur 
Extraterrestrische Physik (MPE), 
National Astronomical Observatories of 
China, New Mexico State University, 
New York University, University of 
Notre Dame, Observat\'ario 
Nacional / MCTI, The Ohio State 
University, Pennsylvania State 
University, Shanghai 
Astronomical Observatory, United 
Kingdom Participation Group, 
Universidad Nacional Aut\'onoma 
de M\'exico, University of Arizona, 
University of Colorado Boulder, 
University of Oxford, University of 
Portsmouth, University of Utah, 
University of Virginia, University 
of Washington, University of 
Wisconsin, Vanderbilt University, 
and Yale University.
\section*{Data Availability}
The data underlying this article are all public and available via the XMM-Newton Science Archive (\hyperlink{https://www.cosmos.esa.int/web/xmm-newton/xsa}{https://www.cosmos.esa.int/web/xmm-newton/xsa}) and the SDSS website (\hyperlink{https://www.sdss4.org}{https://www.sdss4.org/}). The {\sc emcee} samples of the parameters for model (vii) are included in the article's online supplementary material. Additional data products generated for this article will be shared on request to the corresponding author.



\bibliographystyle{mnras}
\bibliography{references} 



\appendix
\section{\hl{Comparison to LaMassa et al. (2016)}}
\label{app:lamassa}
\begin{figure}
    \centering
    \includegraphics[width=\linewidth]{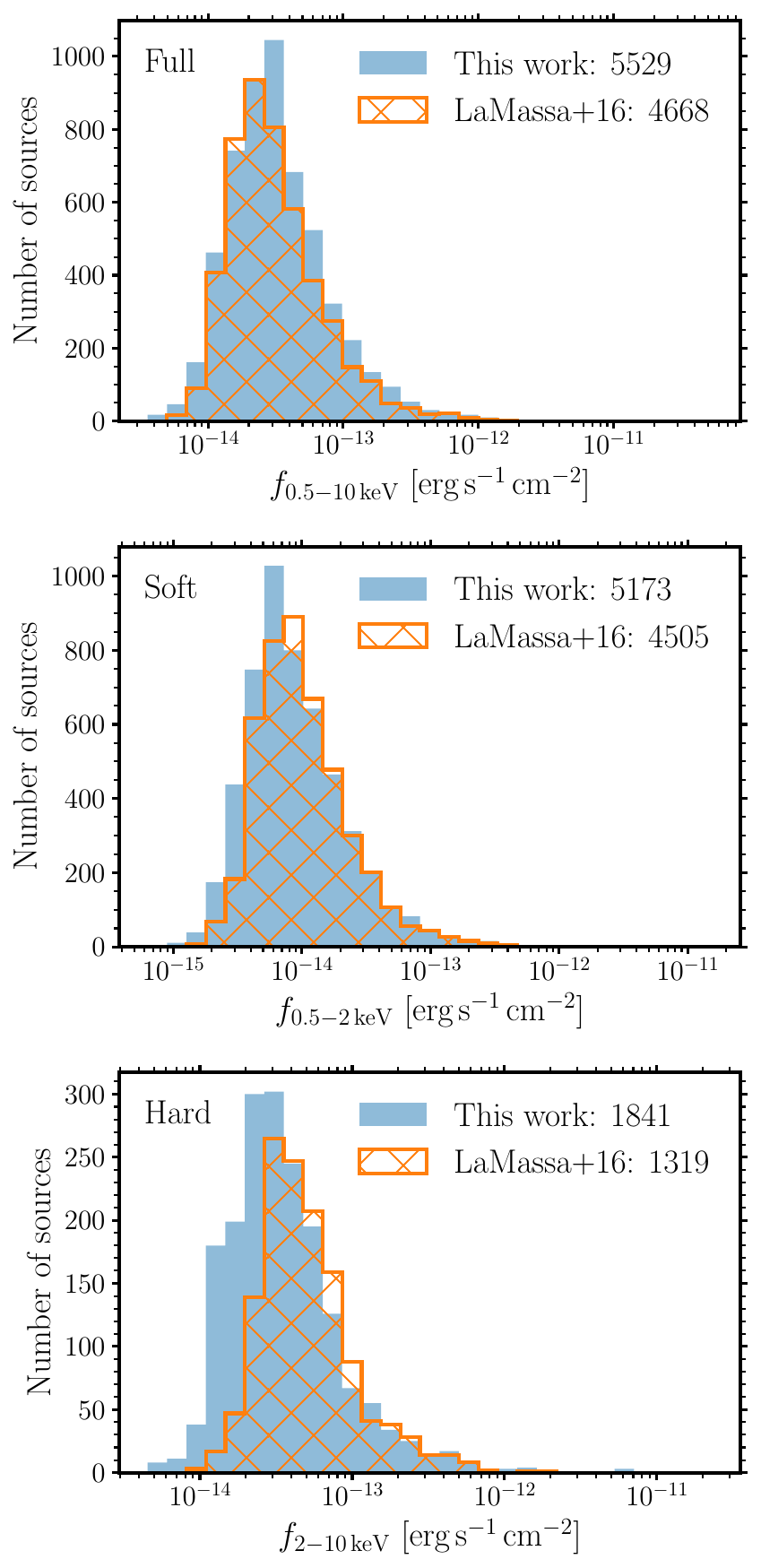}
    \caption{Distribution of source fluxes in the full (top) soft (middle) and hard bands (bottom) for our sample of combined Stripe 82X survey pointings and archival pointings, and the {\lamassa} sample. The number of detected sources in each band are presented in the legends.}
    \label{fig:flux_lam}
\end{figure}

\citet[][hereafter \lamassa]{lamassa_31_2016} have reduced 31\,deg$^2$ of XMM and \textit{Chandra} observations in the Stripe 82 field. In our analysis we have re-reduced the $\sim$28\,deg$^2$ of XMM observations with the {\sc xmmpype} pipeline in order to be consistent with the XXL data, and here we compare our resulting catalogue to that of {\lamassa}. In all three bands (full: 0.5--10\,keV; soft: 0.5--2\,keV; and hard: 2--10\,keV) the {\sc xmmpype} reductions presented in this work produce more sources than in the {\lamassa} sample, an increase of 18\,\%, 15\,\%, and 40\,\% for the full, soft, and hard band detections, respectively. The distributions of fluxes for the three bands in Fig.~\ref{fig:flux_lam} reveal that the majority of the additional detections have low fluxes in the soft and hard bands and there is an increase in bright and faint full-band detections with {\sc xmmpype}. Our source detection algorithm is able to reliably detect fainter sources in at least the soft and hard bands. Note that {\lamassa} converted from counts to fluxes with a photon index of $\Gamma=1.7$ in the full and hard bands, and $\Gamma=2$ in the soft band. We have converted the {\lamassa} fluxes to $\Gamma=1.4$ to match those produced by {\sc xmmpype}. As a reminder, all science was carried out with $\Gamma=1.9$.

\begin{figure}
    \centering
    \includegraphics[width=\linewidth]{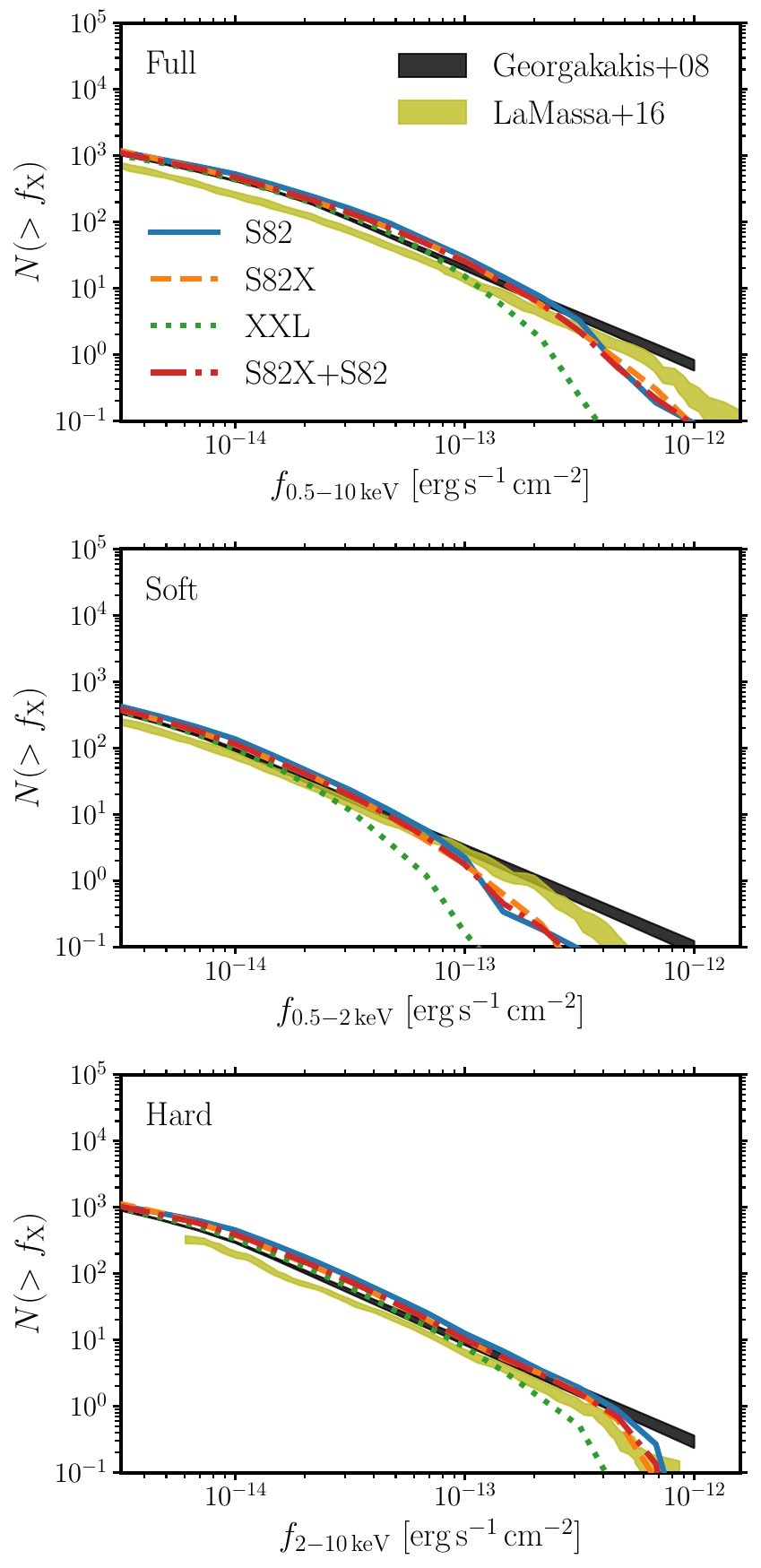}
    \caption{Cumulative number counts as a function of flux for the full band (top), soft band (middle), hard band (bottom). The top panel contains the same information as Fig.~\ref{fig:logNlogS} but we now combine the S82X survey and archival S82 pointings (red) for easy comparison with the {\lamassa} results.}
    \label{fig:lgNlgS_lam}
\end{figure}

We also compare the $\log N$--$\log S$ curves of {\lamassa} and our work in Fig.~\ref{fig:lgNlgS_lam} for the three bands. The combined S82X$+$S82 curves from our work are higher than those of {\lamassa} across all bands at low and intermediate fluxes. In the full and soft bands, our curves are lower at high fluxes, and in the hard band, the high flux end matches well to {\lamassa}; however, the curves are noisier due to fewer sources at the high flux end (see Fig.~\ref{fig:flux_lam}). Encouragingly, our measurements are self-consistent across the three different survey regions and they also agree well with the models of \citet{georgakakis_new_2008} which are based on a range of deeper and higher-resolution \textit{Chandra} surveys.

\section{Derivation of the likelihood}
\label{app:loglike}

We aim to determine the parameters, $\theta$, that maximise the likelihood of observing the data, $\mathcal{D}$, which can be divided into the detected data: $\mathcal{D}_\text{det} = \{N, L_{2500}, z, \text{B}, \text{ECF}, t_\text{exp}\}$; and the undetected data: $\mathcal{D}_\text{not} = \{L_{2500}, z, \mathscr{d}\}$. The detected data contains the number of X-ray source photons, $N$ and the background, ECF and $t_\text{exp}$ at the position of the (X-ray) source. We will group $\{\text{B}, \text{ECF}, t_\text{exp}\}$ as $\mathscr{d}$ for the detected sources. The undetected data contains $\mathscr{d}$ which we are using to describe the overall sensitivity of the surveys, and as such is field-dependent. For brevity, {\lx} and {\luv} are written in place of $\log_{10}L_\text{X}$ and $\log_{10}L_{2500}$. In both cases, $\mathscr{d}$ represents the X-ray data that does not depend on the model or source.

The likelihood can be divided into two terms: the detected data and the undetected data, and we will consider each in turn, starting with the detected data term. The likelihood for detected source $i$, $\mathcal{L}_i$, is the probability of detecting the source and obtaining data $\mathcal{D}_i$ for the model parameters $\theta$:
\begin{equation}
    \mathcal{L}_i = P(\text{det}, \mathcal{D}_i | \theta),
\end{equation}
which can be expanded as,
\begin{equation}
    \mathcal{L}_i = P(\text{det} | N_i, \mathscr{d}_i, L_{2500_i}, z_i, \theta) P(N_i, \mathscr{d}_i, L_{2500_i}, z_i | \theta).
\end{equation}
The first term, $P(\text{det} | N_i, \mathscr{d}_i, L_{2500_i}, z_i, \theta)$, does not depend on $L_{2500_i}$, $z_i$, or $\theta$ and is the probability that an object is detected given that $N_i$ photons were observed. For every one of our detected sources the $N_i$ counts over the background level (contained within $\mathscr{d}_i$) will always satisfy the detection criterion, by definition, thus we have already conditioned on detection and so this term is unity \citep[see ][]{loredo_accounting_2004}. \citet{buchner_obscuration-dependent_2015} summarise why some astronomers still include this term: since typically the luminosity of a source is measured from different data than the data from which the detection was first made (usually with different extraction radii), then this extra step necessitates that the detection probability should remain. In our case, the {\sc xmmpype} reduction uses the same data to determine if a detection meets the detection criteria and then calculates luminosities. The second term can be expanded as follows:
\begin{equation}
    \begin{split}
        P(N_i, \mathscr{d}_i, L_{2500_i}, z_i | \theta) &= P(N_i| \mathscr{d}_i, L_{2500_i}, z_i, \theta) \\
        &\times P(\mathscr{d}_i, L_{2500_i}, z_i | \theta),
    \end{split}
    \label{eq:pNi_theta}
\end{equation}
but as $\mathscr{d}_i$, $L_{2500_i}$, and $z_i$ do not depend on $\theta$ we can drop the second term of equation~\ref{eq:pNi_theta}.

Next, we introduce the marginalisation over {\lx} via, 
\begin{equation}
    \begin{split}
        P(\text{A}) &= \int P(\text{A}, \text{B})\ \text{d}\text{B}\\
        &= \int P(\text{B}|\text{A})P(\text{A})\ \text{d}\text{B},
    \end{split}
\end{equation}
with $B=L_\text{X}$. We have,
\begin{equation}
    \mathcal{L}_i = P(N_i | \mathscr{d}_i,L_{2500_i}, z_i, \theta),
\end{equation}
from which it then follows that,
    \begin{align}
        \mathcal{L}_i &= \int P(N_i, L_\text{X} | \mathscr{d}_i, L_{2500_i}, z_i, \theta)\ \text{d}L_\text{X}\\
        & = \int P(N_i | L_\text{X}, \mathscr{d}_i, L_{2500_i}, z_i, \theta) 
        P(L_\text{X}| \mathscr{d}_i, L_{2500_i}, z_i, \theta)\ \text{d}L_\text{X}.\label{eq:Li_marg2}
    \end{align}
The first term of equation~\ref{eq:Li_marg2} is the probability of observing $N_i$ photons given that for source $i$ we have measured data $\mathscr{d}_i$ and propose that it has an X-ray luminosity {\lx}, which does not depend on $L_{2500_i}$ or $\theta$ and thus reduces to $P(N_i | L_\text{X}, \mathscr{d}_i, z_i)$. This term captures the uncertainty in the value of {\lx} of the source based on the fact that an integer number of counts, $N_i$, were detected; we marginalise over the range of possible {\lx}. With a change of notation, $P(N_i|L_\text{X}, \mathscr{d}_i)$ is equation~\ref{eq:Pni} with $N_\text{exp}$ as the expected number of photons from a source with {\lx} and is calculated via equation~\ref{eq:nexp}. The second term of equation~\ref{eq:Li_marg2} is the prior expectation for {\lx} given the observed $L_{2500_i}$ and $z_i$, and model parameters describing the distribution of {\lx}, independent of the X-ray data (and thus we can drop $\mathscr{d}_i$). The resulting term is the model we aim to fit (equation~\ref{eq:Plx}). Thus, we find
\begin{equation}
    \mathcal{L}_i \propto \int P(N_i|N_{\text{exp}}) P(L_\text{X} | L_{2500_i}, z_i, \theta)\ \text{d}L_\text{X}.
    \label{eq:Li_final}
\end{equation}

Moving onto the likelihood for an undetected source $j$, $\mathcal{L}_j$ is the probability that object $j$ is undetected with data $\mathcal{D}_j$ for the model parameters $\theta$:
\begin{align}
    \mathcal{L}_j &= P(\overline{\text{det}}, \mathcal{D}_j | \theta)\\
    &= P(\overline{\text{det}}, \mathscr{d}_j, L_{2500_j}, z_j | \theta),
\end{align}
which has been marginalised over {\lx}:
\begin{align}
    \mathcal{L}_j &= \int P(\overline{\text{det}}, L_\text{X}, \mathscr{d}_j, L_{2500_j}, z_j | \theta)\ \text{d}L_\text{X} \label{eq:Lj_marg1}\\
    &= \int P(\overline{\text{det}}, L_\text{X} | \mathscr{d}_j, L_{2500_j}, z_j, \theta) P(\mathscr{d}_j, L_{2500_j}, z_j | \theta)\ \text{d}L_\text{X} \label{eq:Lj_marg2}\\
    &= \int P(\overline{\text{det}} | L_\text{X}, \mathscr{d}_j, L_{2500_j}, z_j, \theta) 
    P(L_\text{X} |\mathscr{d}_j, L_{2500_j}, z_j, \theta)\ \text{d}L_\text{X}.\label{eq:Lj_marg3}
\end{align}
The second term in equation~\ref{eq:Lj_marg2} can be dropped for the same reasons as the second term in equation~\ref{eq:pNi_theta}. Equation~\ref{eq:Lj_marg3} results from expanding the first term of equation~\ref{eq:Lj_marg2}.
Equation~\ref{eq:Lj_marg3} is introduced in order to consider the probability of object $j$ remaining undetected for a proposed {\lx}. The first term of equation~\ref{eq:Lj_marg3} can be simplified as $p(\overline{\text{det}} | L_\text{X}, \mathscr{d}_j, z_j)$ and is calculated from the sensitivity curves (see equations~\ref{eq:pndet} and \ref{eq:pdet}) for simplicity we remove $\mathscr{d}_j$ since $\mathscr{d}_j$ is absorbed in the sensitivity curves. The second term reduces to $P(L_\text{X} | L_{2500_j}, z_j, \theta)$ (dropping $\mathscr{d}_j$) and it is again the prior expectation of {\lx} for a given $L_{2500_j}$ and $z_j$. Explicitly,
\begin{equation}
    \mathcal{L}_j \propto \int p(\overline{\text{det}} | L_\text{X}, z_j) P(L_\text{X} |L_{2500_j}, z_j, \theta)\ \text{d}L_\text{X}.
    \label{eq:Lj_final}
\end{equation}
Equations~\ref{eq:Li_final} and \ref{eq:Lj_final} have the same second term (other than different $i$ and $j$ subscripts).

The total likelihood for all of our data is given by the product of $\mathcal{L}_i$ of all detected objects, and $\mathcal{L}_j$ of all undetected object:
\begin{equation}
    \mathcal{L}(\mathcal{D}|\theta) = \prod_i^{N_\text{det}}\mathcal{L}_i \prod_j^{N_\text{not}} \mathcal{L}_j.
\end{equation}
The log-likelihood can therefore be written as,
\begin{equation}
    \begin{split}
        \ln \mathcal{L}(\mathcal{D}|\theta) &= \sum_i^{N_\text{det}} \ln \int P(N_i | N_\text{exp}) P(L_\text{X}|L_{2500_i}, z_i, \theta)\ \text{d}L_\text{X} \\
        &+ \sum_j^{N_\text{not}} \ln \int p(\overline{\text{det}}|L_\text{X}, z_j) P(L_\text{X}|L_{2500_j}, z_j, \theta)\ \text{d}L_\text{X},
    \end{split}
\end{equation}
which is equation~\ref{eq:lnlike}.

\section{\hl{Hard band analysis}}
\label{app:hard}
\begin{figure}
    \centering
    \includegraphics[width=\linewidth]{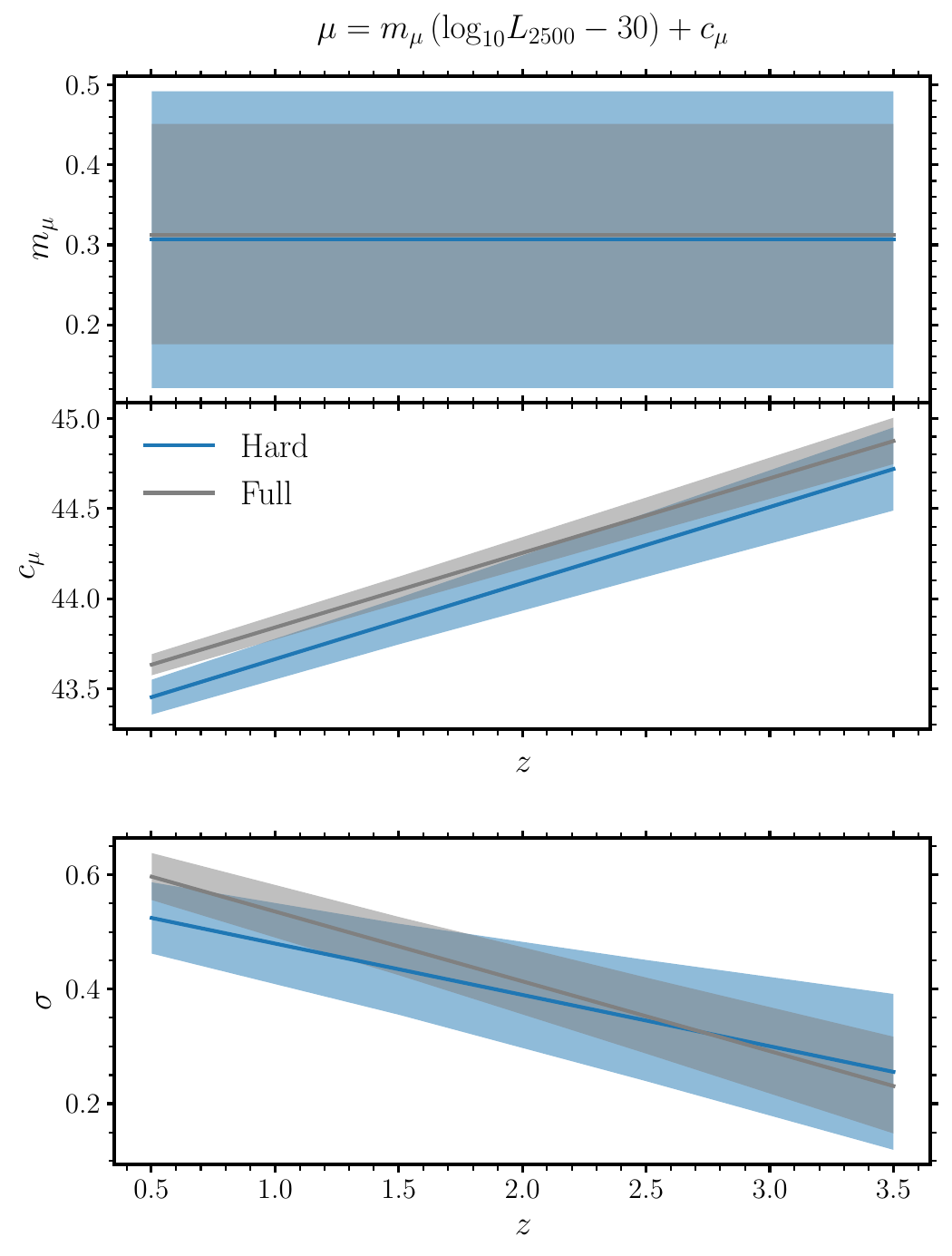}
    \caption{Same as Fig.~\ref{fig:zevo_param} with grey lines and bands corresponding to model (vii) and best-fitting parameters for the full band. The best-fitting hard band model (vii) in blue is consistent with the full band results.}
    \label{fig:params_hard}
\end{figure}

We run our analysis using the same optically-selected sample but with the hard-band X-ray detections and adopting the corresponding sensitivity curve for non-detections. Of the 2292 parent sample, 255 (11\,\%) are detected in the hard band. Due to the much smaller X-ray detected sample, we only run the maximum likelihood estimation for the unbinned models. Table~\ref{tab:AIC_hard} contains the AICs and Multinest evidence for this run. Note that the AIC and relative $\log_{10}Z$ values cannot be compared across different bands as they use different data and have been normalised using their respective best-fit models. The most successful model is model (vii) which is the same as for the full band. Models (vi) and (viii) also prove to fit the data well with the former being preferred by the Multinest Bayesian evidence and the latter by the AIC. Importantly, all three of these models allow redshift evolution. In Fig.~\ref{fig:params_hard} the model parameters for model (vii) using the hard band data are consistent with those for the full band data in the main paper.
The consistency between our hard-band results, presented here, and the full-band analysis used in the main paper demonstrates both the robustness of our Bayesian analysis when applied to smaller, shallower samples (as is the case for the hard-band sample) and that X-ray absorption effects are not driving these results and our observed redshift evolution (as any impact would be severely reduced when using a harder band). 

\begin{table*}
    \centering
    \caption{Same as Table~\ref{tab:AIC} but for the hard-band sample and only the unbinned models.}
    \begin{tabular}{c|c|c|c|c|c|c}
\hline
Model & Binning & Parameters & $N_{\text{dim}}$ & AIC & $\Delta$AIC & $\log_{10} Z$\\
\hline
 (iv) & Unbinned, no $z$ evolution & $m_\mu$, $c_\mu$, $\sigma$ & 3 & 3435.27 & 41.82 & -7.1\\
 (v) & Unbinned, no $z$ evolution & $m_\mu$, $c_\mu$, $m_\sigma$, $c_\sigma$ & 4 & 3426.67 & 33.23 & -7.1\\
 (vi) & Unbinned, $z$ evolution & $m_\mu(z)$, $c_\mu(z)$, $\sigma$ & 5 & 3400.37 & 6.92 & -1.3\\
 \textbf{(vii)} & \textbf{Unbinned, \bm{$z$} evolution} & \textbf{\bm{$m_\mu$}, \bm{$c_\mu(z)$}, \bm{$\sigma(z)$}} & \textbf{5} & \textbf{3393.45} & \textbf{0.00} & \textbf{0.0}\\
 (viii) & Unbinned, $z$ evolution & $m_\mu(z)$, $c_\mu(z)$, $\sigma(z)$ & 6 & 3395.16 & 1.71 & -2.2\\
\hline

    \end{tabular}
    \label{tab:AIC_hard}
\end{table*}

\section[Derivation of \texorpdfstring{\aox}{aox} relation]{Derivation of $\vect{\alpha_{\text{ox}}}$ relation}
\label{app:aoxr}
By combining model (vii) and equation~\ref{eq:aox} we derive the relationship between the peak of the distribution of {\aox} and the {\luv} and redshift.
The peak of the {\lx} distribution, in units of erg\,s$^{-1}$, is given by $\mu=m_\mu\left({\log_{10}}L_{2500} - 30\right) + p_{\mu}z + k_{\mu}$. The monochromatic 2\,keV luminosity is estimated from the full-band (0.5--10\,keV) luminosity via
\begin{equation}
    L_{E} = \frac{(2-\Gamma)L_{0.5-10\,\text{keV}}}{10\,\text{keV}^{2-\Gamma}-0.5\,\text{keV}^{2-\Gamma}}E^{1-\Gamma}
    \label{eq:monolum}
\end{equation}
in units of erg\,s$^{-1}$\,keV$^{-1}$ with $E=2$\,keV and $\Gamma=1.9$.
Substituting $\mu$ in for $L_{0.5-10\,\text{keV}}$ and multiplying the resulting monochromatic luminosity by a factor h ($=4.136\times10^{-18}$\,keV\,Hz$^{-1}$) produces $L_{2\,\text{keV}}$ in units of erg\,s$^{-1}$\,Hz$^{-1}$.

Multiplying equation~\ref{eq:monolum} by h and inserting into equation~\ref{eq:aox} generates the following,
\begin{equation}
    \begin{split}
        &\alpha_{\text{ox}}(L_{2500}, z) = -6.9727 - 0.3838\log_{10}\left(\frac{L_{2500}}{{\text{erg}}\,{\text{s}}^{-1}\,{\text{Hz}}^{-1}}\right)\\
        &+ 0.3838\left[m_\mu\left(\log_{10}\left(\frac{L_{2500}}{{\text{erg}}\,{\text{s}}^{-1}\,{\text{Hz}}^{-1}}\right) - 30\right) \right.
        + \left. p_{\mu}z + k_{\mu}\vphantom{\left(\frac{L_{2500}}{{\text{erg}}\,{\text{s}}^{-1}\,{\text{Hz}}^{-1}}\right)}\right],
    \end{split}
    \label{eq:aoxr_full}
\end{equation}
where the constant $-6.9727$ encompasses the constant values from $L_{2\,\text{keV}}$ and the factor of 0.3838 which is the denominator of equation~\ref{eq:aox}. Gathering all {\luv} terms, all $z$ terms and all constants, one arrives at 
\begin{align}
    &a\log_{10}L_{2500} = 0.3838\left(m_\mu - 1\right)\log_{10}L_{2500}, \\
    &bz = 0.3838 p_{\mu}z,
\end{align}
and,
\begin{equation}
    c = -6.9727 - 0.3838\left(30m_\mu + k_{\mu}\right),
\end{equation}
ultimately arriving at
\begin{equation}
    \alpha_{\text{ox}}(L_{2500}, z) = a \log_{10}\left(\frac{L_{2500}}{{\text{erg}}\,{\text{s}}^{-1}\,{\text{Hz}}^{-1}}\right) + b z + c,
\end{equation}
which is equation~\ref{eq:aoxr}.
\begin{figure*}
    \centering
    \includegraphics[width=0.8\linewidth]{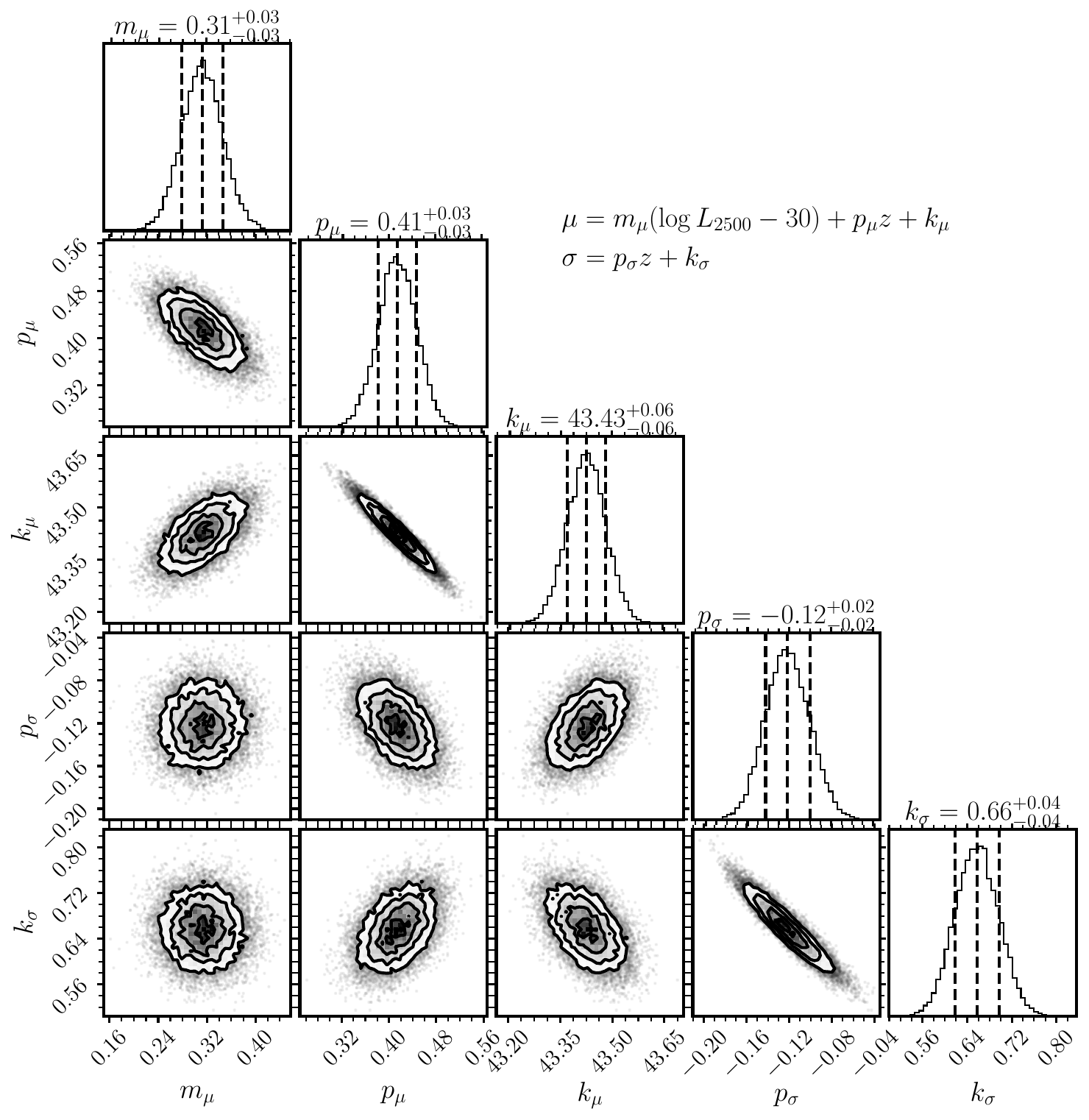}
    \caption{Posterior distributions of the model parameters. Some parameters show correlations, namely $p_{\mu}$ with $k_{\mu}$, and $p_{\sigma}$ with $k_{\sigma}$.}
    \label{fig:corner}
\end{figure*}

The uncertainty on {\aox} is given by
\begin{equation}
    \begin{split}
        \Delta\alpha_{\text{ox}} = \left[\vphantom{\Delta c^2}\right.&({\log}L_{2500}\Delta a)^2 + (z\Delta b)^2 + \Delta c^2 \\
        &+ 2{\log}L_{2500}\,\text{cov}[a, c] + 2z\,\text{cov}[b, c] \\
        &+ \left.\vphantom{\Delta c^2} 2{\log}L_{2500}\,z\,\text{cov}[a,b]\right]^{1/2}
    \end{split}
\end{equation}
where cov[X, Y] is the covariance between parameters X and Y. For brevity, $\log_{10}L_{2500}$ in erg\,s$^{-1}$\,Hz$^{-1}$ is represented by $\log L_{2500}$. The various covariances are calculated as follows:
\begin{align}
    &\text{cov}[a,c] = 0.3838^2\left(\text{cov}[m_\mu, k_{\mu}] - 30\Delta m_\mu^2\right), \\
    &\text{cov}[b, c] = 0.3838^2\left(\text{cov}[p_{\mu}, k_{\mu}] - 30\,\text{cov}[p_{\mu}, m_\mu]\right), \\
    &\text{cov}[a, b] = 0.3838^2\,\text{cov}[m_\mu, p_{\mu}].
\end{align}
We provide the {\sc emcee} samples of the parameters for model (vii) as supplementary data to allow calculation of the covariances and uncertainties. We note that we have assumed that the model parameters are independent; however, in Fig.~\ref{fig:corner} we see strong correlation between $p_{\mu}$ and $k_{\mu}$ which is to be expected since these parameters describe the linear relationship between redshift and $\mu$. The same can be said for $p_{\sigma}$ and $k_{\sigma}$, the gradient and intercept of $\sigma(z)$ (although they do not enter into the equation for {\aox}).


\bsp	
\label{lastpage}
\end{document}